\documentstyle[preprint,aps,eqsecnum]{revtex}

\begin{document}

%%\;\raisebox{-.4ex}{\rlap{$\sim$}} \raisebox{.4ex}{$<$}\;
%%\;\raisebox{-.4ex}{\rlap{$\sim$}} \raisebox{.4ex}{$>$}\;

{\tighten
\preprint{\vbox{\hbox{FERMILAB--PUB--94/361--T}}}

\title{$B_s -\overline{B}_s$ Mixing, CP Violation and Extraction of CKM
Phases from Untagged $B_s$ Data Samples}

\author{Isard Dunietz}
\address{Fermi National Accelerator Laboratory \\ P.O.~Box 500,
Batavia, Illinois 60510}

\bigskip
\date{\today}

\maketitle
\begin{abstract}
A width difference of the order of 20\% has previously been predicted for the
two mass eigenstates of the $B_s$ meson.  The dominant contributor to the width
difference is the
$b\rightarrow c\bar c s$ transition, with final states common to both
$B_s$ and $\overline{B}_s$. All current experimental analyses fit the
time-dependences of flavor-specific $B_s$-modes to a single
exponential, which essentially determines the average $B_s$ lifetime.  We
stress that the same data sample allows even the measurement of the width
difference. To see that, this note reviews the time-dependent formulae
for tagged $B_s$ decays, which involve rapid oscillatory terms depending
on $\Delta mt$.  In untagged data samples the rapid oscillatory terms cancel.
Their time-evolutions depend only on the much more slowly
varying exponential falloffs.  We discuss in detail the
extraction of the two widths, and identify the large (small) CP-even (-odd)
rate
with that of the light (heavy) $B_s$ mass eigenstate.
It is demonstrated that decay length distributions of some \underline{untagged}
$B_s$ modes, such as $\rho^0 K_S, \; D_s^{(*)\pm}K^{(*)\mp}$, can be used to
extract the notoriously difficult CKM unitarity triangle angle $\gamma$.
Sizable CP violating effects may be seen with such untagged $B_s$ data samples.
 Listing $\Delta\Gamma$ as an observable allows for additional important
standard model constraints.  Within the CKM model, the ratio $\Delta\Gamma/
\Delta m$ involves no CKM parameters, only a QCD uncertainty.
Thus a measurement
of $\Delta\Gamma \;(\Delta m)$ would predict $\Delta m \;(\Delta\Gamma )$, up
to the QCD uncertainty.  A large width difference would automatically solve
the puzzle of the number of charmed hadrons per $B$ decay in favor of theory.
We also derive an upper limit of $(| \Delta\Gamma | / \Gamma)_{B_s}
\;\raisebox{-.4ex}{\rlap{$\sim$}} \raisebox{.4ex}{$<$}\;0.3$.  Further,
we must abandon the notion of branching fractions of
$B_s\rightarrow f$, and instead consider
$ B(B^0_{L(H)}\rightarrow f)$,
in analogy to the neutral kaons.
\end{abstract}
\pacs{}
}% end the tighten

\section{Introduction}

$B$ physics has matured to the point that data samples of strange
$B$ mesons are currently being collected both at Fermilab \cite{bscdf} and at
LEP \cite{bslep,opal,aleph,delphi}.
More than 200 flavor-specific events and a few dozen
$J/\psi\phi$ events have been recorded. It is believed that precision studies
of $B_s$
mesons requires a distinction between $B_s$ and $\overline{B}_s$ mesons
(henceforth denoted as ``tagging") and superb vertex resolution so as to follow
the rapid oscillatory behavior dependent upon $\Delta mt$. Then the observation
of
CP-violating phenomena and the extraction of fundamental (Cabibbo-Kobayashi
-Maskawa~\cite{ckm}) CKM-parameters can
be contemplated~\cite{paschoszacher,dr}.

It may not be imperative to trace the rapid $\Delta mt$-oscillations.
Time-dependent studies of \underline{untagged} data samples of $B_s$'s
remove the rapid oscillatory behavior depending upon $\Delta mt$.
What remains are
two exponents $e^{-\Gamma_L t}$ and $e^{-\Gamma_H t}$, where the
light and heavy $B_s$-mass eigenstates have an average lifetime of about
$\tau_b \sim 1.6 \;ps$~\cite{forty}, and are expected to differ by about
(20-30)\% ~\cite{box,shifman,dpt,dthesis,dunietzwidth,aleksan,bigi,bigietal}.
This could be sufficient for observation of $B_s -\overline B _s$ mixing
(due to lifetime differences), CP-violation and the clean extraction
of CKM-parameters.  Tagging and time-resolving $\Delta mt$ oscillations would
of course allow
many additional precision $B_s$-measurements (for reviews see for instance
Refs.~\cite{cpreviewBigi,cpreviewNir,cpreviewMcDonald}).

Lately there has been an emphasis on the predicted large
mass-mixing,
\begin{equation}
x_s =\left(\frac{\Delta m}{\Gamma}\right)_{B_s} \;.
\end{equation}
The measurement of $x_s$ requires tagged $B_s$ data samples and superb vertex
resolution for tracing the rapid $\Delta mt$ oscillations~\cite{xs}.  The
parameter $x_s$ may turn out to be too large to be measured in the foreseeable
future~\cite{xs,xsnir,sharma}.   There exists, however, another clear measure
of $B_s -\overline{B}_s$ mixing,
namely, a width difference $\Delta\Gamma$ between the $B_s$ mass-eigenstates.
The ratio $\Delta\Gamma / \Delta m$ has been estimated~\cite{dpt,dthesis}.  It
suffers from no CKM-uncertainty only from hadronic uncertainties.  Thus, large
$\Delta m$-values that are currently impossible to measure may imply values for
$\Delta\Gamma$ that are currently feasible.
It may happen that a width difference will be the first observed $B_s -
\overline{B}_s$ mixing effect.

The implications of measuring a non-zero $\Delta\Gamma$
would be far reaching. Not only would $B_s -\bar B_s$ mixing
be demonstrated, but $\Delta m$ would perhaps be well estimated.
The estimate would combine the predicted ratio $\Delta m/\Delta\Gamma$
with the more traditional approaches~\cite{xsnir} to optimize our
knowledge on $\Delta m$.
A reliable estimate or measurement of $\Delta m$ allows not only the
extraction of the combination of decay constant and bag parameter$\;\;
(B_{B_s}f^2_{B_s})$ \cite{box}, but even
the planning of a multitude of CP-violating measurements and determinations of
CKM-parameters with \underline{tagged} $B_s$-data samples
\cite{cpreviewBigi,cpreviewNir,cpreviewMcDonald}.
(Conversely, if $\Delta m$ were to be observed first, valuable
information on $\Delta\Gamma$ would be available. In the long term,
measurements of both $\Delta m$ and $\Delta\Gamma$ allows us to probe
the hadronic uncertainties arising in $\Delta\Gamma /\Delta m$.)
Some of the central points of this note follow. First, a non-vanishing
$\Delta\Gamma$ enables us to observe large CP-violating effects and to cleanly
extract CKM-parameters (for instance $\gamma$) from much more slowly varying
time-evolutions of some \underline{untagged} $B_s$-data samples.

In contrast, the traditional methods that use $B_s$-decays require
tagging and the ability to trace the rapid $\Delta mt$-oscillations.
It is easy to explain why such measurements are possible for non-zero
$\Delta\Gamma$ with some untagged data samples.
Consider the creation of a $B_s$. The $B_s$ state can be written as
a linear superposition of the heavy $B_H$ and light $B_L$
eigenstates of the mass matrix. Because the two eigenstates have different
lifetimes, suitably long times can be chosen where
the longer lived $B_H$ is highly enriched, $|B_H \rangle  = p|B_s \rangle \;
-\; q|\overline B_s \rangle $.
Time is the tag here, in analogy to the neutral kaons.

Consider now any $B_s$-mode $f$ that can be fed from both a $B_s$
and $\overline B_s$, and where the two unmixed amplitudes
$(\langle f|B_s \rangle $ and $\langle f|\overline B_s\rangle )$ differ in
their CKM-phase.
Those modes then could harbor observable CP-violating effects. Further,
it will become clear (by the end of this note) how to determine the
CKM-phase difference.
For instance, the CKM-angle $\gamma$ can be determined from the
\underline{untagged} $\rho^0 K_S$ data sample if
penguin amplitudes are negligible.
Penguin diagrams may be sizable, in which case $\gamma$ can be determined
from untagged $D^{(*)\mp}_s K^{(*)\pm}$ data samples.
This last determination assumes factorization for the color-allowed processes
$B_s\rightarrow D^{(*)-}_s K^{(*)+} ,\; D^{(*)-}_s \pi^+ $.

To those who object to this factorization assumption, we offer
the extraction of $\gamma$ without any theoretical input from the
\underline{untagged} $D^0 \phi ,\overline D^0 \phi$ and
$D^0_{CP}\phi$ data samples.
$D^0_{CP}$ denotes that the
$D^0$ or $\overline{D^0}$ is seen in a CP-eigenmode, such as
$\pi^0 K_S ,\;K^+K^- ,\;\pi^+\pi^-$.
Clearly all those above-mentioned processes (and many more) could show sizable
CP violating effects, which we discuss.

Second, a large width difference would solve rather convincingly
the charm deficit puzzle in favor of theory~\cite{baffle,fwd,voloshin,bagan},
because $B(b\rightarrow c\bar c s )\;\raisebox{-.4ex}{\rlap{$\sim$}}
\raisebox{.4ex}{$>$}\;
\left( |\Delta\Gamma| / \Gamma \right)_{B_s}$.
Third, if hadronic effects could be controlled and understood, $f_{B_s}$ could
be
extracted from a measurement of $\Delta\Gamma$.
Fourth, one would not be allowed to speak about
branching fractions of an unmixed $B_s$ to any final state $f$, but rather
one would have to discuss $B(B_{H(L)} \rightarrow f$).

The derivation of a reliable upper limit for $|\Delta\Gamma |/\Gamma
\;\raisebox{-.4ex}{\rlap{$\sim$}} \raisebox{.4ex}{$<$}\; 0.3$ is also of some
importance, because it informs us about the optimal size of such effects.
Establishing a non-vanishing width difference is thus important, because of all
the above-mentioned reasons.

Bigi et al. suggested the use of the $J/\psi \phi$ and $D_s^\mp \ell^\pm \nu$
data samples to extract the width difference \cite{bigi,bigietal}.  This note
reviews and refines that suggestion and discusses other determinations of
$\Delta\Gamma$.  What is intriguing is that $\Delta\Gamma$ could be measured
from currently available data samples with more statistics, which are the
\underline{untagged}, flavor-specific modes of $B_s$.  Such $B_s$ modes
time-evolve as the sum of two exponentials~\cite{dpt,dr},
\begin{equation}
e^{-\left(\Gamma + \frac{\Delta\Gamma}{2}\right)t} + e^{-\left(\Gamma -
\frac{\Delta\Gamma}{2}\right)t} \;.
\end{equation}
A one parameter fit for $\Delta\Gamma$ determines the width difference.  The
average width $\Gamma$ of $B_s$ is well known.  It can be obtained essentially
from a one parameter fit of the time-evolution of that same (untagged,
flavor-specific $B_s$) data sample to a single exponential
$\exp(-\Gamma t)$~\cite{feasibility}.  Alternatively one can either use the
prediction that $\Gamma$ equals the $B_d$ width to sufficient
accuracy~\cite{bigi,bigietal}, or one can obtain $\Gamma$ from the average
$b$-hadron lifetime determined in high energy experiments.  Several additional
methods for extracting the width difference will become available in the
future.  This note discusses a few of them.  A careful feasibility study will
be reported elsewhere~\cite{feasibility}.

This report is organized as follows. Section II
reviews $B_s -\overline{B}_s$ mixing
phenomena. Section III lists a few
ramifications of a sizable difference in widths, and derives an upper limit of
$(| \Delta\Gamma | / \Gamma)_{B_s} \;\raisebox{-.4ex}{\rlap{$\sim$}}
\raisebox{.4ex}{$<$}\;0.3$.   Section IV discusses time-evolution of $B_s$
mesons and finds that any rapid oscillatory behavior depending on
$\Delta mt$ cancels in untagged data samples. Suggestions for the experimental
determination of $\Delta\Gamma$, CP-violation, and CKM-parameters with
untagged $B_s$ samples can be found in Section V. Section VI concludes.

\section{Predictions for $B_s -\overline{B}_s$ Mixing}

This section collects a few pertinent mixing formulae from the general
treatment reviewed in Chapter 5 of Ref. \cite{dthesis}.
An arbitrary neutral $B_s$-meson state
\begin{equation}
a \mid B_s\rangle + b\mid \overline{B}_s\rangle
\end{equation}
is governed by the time-dependent Schr\"odinger equation
\begin{equation}
i \;\frac{d}{dt} \;\left(\begin{array}{c} a \\ b \end{array}
\right) = {\bf H}
\left(\begin{array}{c} a \\ b \end{array}
\right) \equiv \left({\bf M} -\frac{i}{2} {\bf \Gamma}\right) \;
\left(\begin{array}{c} a \\ b \end{array}\right) \;.
\end{equation}
Here ${\bf M}$ and ${\bf \Gamma}$ are $2\times 2$ matrices, with ${\bf M}=
{\bf M}^+ , {\bf \Gamma} = {\bf \Gamma}^{+}$. CPT invariance guarantees
$M_{11} = M_{22}$ and $\Gamma_{11} = \Gamma_{22}$. We assume
CPT throughout and obtain the eigenstates of the mass matrix as
\begin{equation}
\mid B_L\rangle=p\mid B^0_s\rangle + \;q\mid \overline{B}^0_s\rangle \;,
\end{equation}

\begin{equation}
\mid B_H\rangle=p\mid B^0_s\rangle - \;q\mid\overline{B}^0_s\rangle \;,
\end{equation}
with eigenvalues ($L=$ ``light", $H= $``heavy")
\begin{equation}
\mu_{L,H} =m_{L,H} -\frac{i}{2} \;\Gamma_{L,H} \;.
\end{equation}
Here $m_{L,H}$ and $\Gamma_{L,H}$ denote the masses and decay widths
of $B_{L,H}$. Further, define
\begin{equation}
\Delta\mu \equiv\mu_H -\mu_L \equiv\Delta m-\frac{i}{2} \;\Delta\Gamma \;, \;\;
\Gamma \equiv\frac{\Gamma_L +\Gamma_H}{2} \;.
\end{equation}
Within the CKM model, the dispersive $M_{12}$ and absorptive $\Gamma_{12}$
mass matrix elements satisfy \cite{box,dthesis}
\begin{equation}
\mid M_{12} \mid \;\gg \;\mid\Gamma_{12} \mid \;,
\end{equation}
and thus~\cite{box,dthesis}
\begin{equation}
\Delta m \approx 2 \mid M_{12} \mid \;.
\end{equation}
$M_{12}$ is by far dominated by the virtual $t\bar t$ intermediate state
and
\begin{equation}
M_{12} \approx - c \;\xi^2_t \;,
\end{equation}
Here
\begin{equation}
\xi_q = V_{qb} \;V^*_{qs}
\end{equation}
and $c$ is a positive quantity under the phase convention
\begin{equation}
CP \mid B_s\rangle =+\mid\overline{B}_s\rangle \;.
\end{equation}
The coefficients $q/p$ satisfy
\begin{equation}
\frac{q}{p} =\frac{-\Delta \mu}{2(M_{12} -\frac{i}{2} \Gamma_{12})} \;.
\end{equation}
The CKM model predicts
\begin{equation}
\left|\frac{q}{p}\right| =1+{\cal O} (10^{-3} - 10^{-4}) \;.
\end{equation}
The width difference is precisely \cite{dthesis}
\begin{equation}
\Delta\Gamma =\frac{4 \;Re \;(M_{12} \Gamma^*_{12})}{\Delta m} \;.
\end{equation}

Modes that are common to $B_s$ and $\overline{B}_s$ contribute to $\Gamma_{12}$
and thus determine $\Delta\Gamma$, see Eq.~(2.14).
The most dominant modes are governed
by the CKM-favored $b\rightarrow c\bar c s$ transition, with the CKM-suppressed
$b\rightarrow c\bar u s, u\bar c s, u\bar u s$ processes playing a
minor role \cite{box}.

Box diagram calculations \cite{box,shifman,aleksan} yield a negative
$\Delta\Gamma$,
\begin{equation}
\frac{\Delta\Gamma}{\Gamma} \sim (-0.2)\;.
\end{equation}
In addition, Ref.~\cite{aleksan} employed an orthogonal approach of
summing over many exclusive modes governed by the $b\rightarrow c \bar c s$
process.  Denote by $\Gamma_+ (b\rightarrow c\bar c s) \;[\Gamma_-
(b\rightarrow c\bar c s)]$
the CP-even [CP-odd] rate governed
by the $b\rightarrow c\bar c s$ transition of the $B_s$ meson.
Ref.~\cite{aleksan} finds that $\Gamma_+ (b\rightarrow c\bar c s)$ by far
dominates $\Gamma_- (b\rightarrow c\bar c s)$, and again a width difference
of $\sim$ 20\% results,
\begin{eqnarray}
\Gamma_+ (b\rightarrow c\bar c s) \; & \gg & \;\Gamma_- (b\rightarrow
c\bar c s), \;\frac{\Gamma_+ (b\rightarrow c\bar c s) -\Gamma_-
(b\rightarrow c\bar c s)}{\Gamma_+ (b\rightarrow c\bar c s) +\Gamma_-
(b\rightarrow c\bar c s)}
=0.97 ,\nonumber \\
& & \frac{\Gamma_+ (b\rightarrow c\bar c s) -\Gamma_- (b\rightarrow
c\bar c s)}{\Gamma} \sim 0.2 \;.
\end{eqnarray}
The significant fraction of baryonic modes, such as $B_s\rightarrow
\Xi^{(r)}_c \overline{\Xi}^{(r)}_c$, was not considered, however.

CP violating effects of $B_s$ decays governed by the
$b\rightarrow c\bar c s$ transition are tiny.
Neglecting CP violation, the heavy and light mass-eigenstates also have
definite CP properties,~\cite{confusion}
\begin{equation}
\Gamma_H =\Gamma_- ,\;\;\Gamma_L =\Gamma_+ \;.
\end{equation}
The identification [Eq. (2.17)] will be seen from yet another viewpoint later
on in Section (V.B). The box diagram calculation and the orthogonal approach of
summing over many exclusive modes both predict the same sign for
$\Delta\Gamma$.

\section{Consequences of sizable $(\Delta \Gamma)_{B_s}$}

A large width difference $\Delta\Gamma$ would have important implications
for several areas of the Standard Model. We discuss only a few consequences
such
a $\Delta\Gamma$ measurement would make.
First, within the CKM-model the ratio $\frac{\Delta m}{\Delta
\Gamma}$ can be estimated~\cite{dpt,dthesis},
\begin{equation}
\frac{\Delta m}{\Delta\Gamma} \approx \frac{-2}{3\pi}
\;\;\frac{m^2_t\;h(m^2_t/M_W^2)}{m^2_b}
\;\;\left(1-\frac{8}{3} \;\frac{m^2_c}{m^2_b} \right)^{-1} \;,
\end{equation}
where\cite{inamiLim}
\begin{equation}
 h(y) =  1 - \frac{3\;y\;(1+y)}{4\;(1-y)^2}
\;\bigg\{1\;+\;\frac{2\;y}{1-y^2} \;\ln(y) \bigg\} \;.
\end{equation}
The quantity $\Delta m / \Delta\Gamma$ has no CKM ratio.  In contrast, the
correction to Eq.~(3.1) involves a QCD uncertainty.
It is imperative to estimate sensibly the error upon such a QCD based
calculation.
If the error does not turn out to be too large, then a measured $\Delta\Gamma$
implies an
allowed range for $\Delta m$, or vice-versa (depending upon which measurement
comes first). If the ratio $\Delta m / \Delta\Gamma$ could be reliably
calculated, then $|V_{td} / V_{ts}|^2$ could be determined by combining the
measurement of $(\Delta\Gamma)_{B_s}$ with the $B_d - \overline{B}_d$ mixing
parameter $(\Delta m)_{B_d}$~\cite{browderpakvasa}.  The ratio $(\Delta m /
\Delta\Gamma)_{B_s}$ could become another Standard Model constraint.

Second, we have previously shown how to extract angles of the
unitarity CKM triangle from time-dependent studies of $B_s$ and/or
$B_d$ \cite{dsnowmass}, assuming a vanishing width difference.
If a non-zero $(\Delta\Gamma )_{B_s}$ were to be found, those studies would
have to be
modified. We are confident that the angles of the unitarity
CKM triangle can still be extracted from those correlations.
The demonstration of this fact goes beyond the scope of this report, however.

Third, a large width difference would solve the so-called
puzzle of the number of charmed hadrons per $B$-meson $n_c$, which we will
demonstrate. Theoretically
we expect $n_c \;\approx \; 1.3$
\cite{baffle,fwd,voloshin,bagan}, whereas the current world average is $1.11\;
\pm \; 0.06$~\cite{muheim}.  Frankly, we do not perceive the apparent
discrepancy as a problem.  After scrutinizing the experimental data, we
realized that the uncertainties in the branching
fractions of the decays of the more exotic charmed hadrons could be
under-estimated. Also, the detection efficiencies of the more exotic charmed
hadron species in $B$ decays have yet to be carefully analyzed.
It is possible that experiments will eventually agree with theory,
$n_c \;\approx \; 1.3$.
However, a large $(-\Delta\Gamma /\Gamma )_{B_s}$ would give direct proof
that $B(b\rightarrow c\bar c s)$ is large (here we neglect the tiny
$W$-annihilation amplitude
$b \bar s \to c \bar c$ and the
small corrections that must be incorporated now that widths of the heavy
and light
$B_s$  differ), because
\begin{equation}
B(b\rightarrow c\bar c s )\;\raisebox{-.4ex}{\rlap{$\sim$}}
\raisebox{.4ex}{$>$}\;
\left(\frac{-\Delta\Gamma}{\Gamma}\right)_{B_s} \;.
\end{equation}
Eq.~(3.3) follows from the following steps
\begin{eqnarray}
B(b\rightarrow c\bar c s) & = & \frac{\Gamma (b\rightarrow c\bar c s)}{\Gamma}
=
\nonumber \\
& = & \frac{\Gamma_+ (b\rightarrow c\bar c s)+\Gamma_- (b\rightarrow
c\bar c s)}{\Gamma} \geq \frac{\Gamma_+ (b\rightarrow c\bar c s) -\Gamma_-
(b\rightarrow c\bar c s )}{\Gamma} \approx \nonumber \\
& \approx & \frac{-\Delta\Gamma}{\Gamma} \;,
\end{eqnarray}
where $\Gamma_+ (b\rightarrow c\bar c s) \;[\Gamma_- (b\rightarrow c\bar c s)]$
denotes the CP-even [CP-odd] width of the $B_s$ modes governed
by the one dominant CKM-favored $b\rightarrow c\bar c s$ transition.  The
inclusive width of $B_s$ mesons governed by the $b\rightarrow c\bar c s$
process is denoted by $\Gamma (b\rightarrow c\bar c s)$ and
satisfies~\cite{lipkin}
\begin{equation}
\Gamma (b\rightarrow c\bar c s) = \Gamma_+ (b\rightarrow c\bar c s) + \Gamma_-
(b\rightarrow c\bar c s)\; .
\end{equation}
This equation was used in the second step of Eq.~(3.4).
Thus a large width difference $\Delta\Gamma$ implies directly a large
branching fraction for the $b\rightarrow c\bar c s$ transition, see Eq.~(3.3).

QCD calculations in $b$ decays have progressed far enough that a reliable upper
limit for $(| \Delta \Gamma | /\Gamma)_{B_s}$ can be obtained.  The least
trustworthy QCD estimate is that for $\Gamma(b\rightarrow c\bar c s)$, because
the sum of the masses of the three final quarks are at the $m_b$-scale.
Uncalculable non-perturbative and resonant effects may be important. This is
borne out from data at $\Upsilon(4S) \to B \overline B$, where the $D_s$
momentum spectrum indicates that about half of all the $D_s$ in $B$ decays
originate from two-body $B$-modes~\cite{Dscleo}.  Thus a QCD-corrected parton
calculation may not be quantitatively applicable to $\Gamma(b\rightarrow c\bar
c s)$.  However the width for $b\rightarrow c\bar c s$ can be obtained
indirectly~\cite{fwd},

\begin{eqnarray}
B(b\rightarrow c\bar c s) & \approx &
|V_{cs}|^2\;(1\;-\;\sum_{\ell}\;B(b\rightarrow c\ell \nu) - B(b\rightarrow
c\bar u d'))\;= \nonumber \\
& = & |V_{cs}|^2\;(1\;-\;\sum_{\ell}\;B(b\rightarrow c\ell \nu) -
\frac{\Gamma(b\rightarrow c\bar u d')}{\Gamma(b\rightarrow c e \nu)}\;
B(b\rightarrow c e \nu))\;,
\end{eqnarray}
$$|V_{cs}|^2 \; \approx \; 1\;-\;\theta^2_c.$$
We neglect rare processes, such as those mediated by an underlying $b \to u$
transition or penguin induced decays.  Here $d'$ and $s'$ denote the weak
eigenstates ($d' = d \cos \theta_c
- s \sin \theta_c,
s' = d \sin \theta_c + s \cos \theta_c$) and $\sin
\theta_c\;\approx\;\theta_c\;\approx\;0.22$ is the
Cabibbo angle.  The highly involved $\alpha_s$ corrections for the
$b\rightarrow c\bar u d'$ rate for a massive charm have been completed recently
by Bagan
et al.~\cite{bcudbagan}; see also earlier work~\cite{earlier}.  The ratio

\begin{equation}
\frac{\Gamma (b\rightarrow c\bar ud')}{\Gamma (b\rightarrow ce\nu )}
= 3\; \eta_{QCD} \approx 3 \cdot 1.35
\end{equation}
is thus well known theoretically~\cite{bagan}, and the semileptonic
branching fractions have been measured~\cite{pdg,slBR}

\begin{equation}
B(B\rightarrow Xe\nu ) = (10.7\;\pm \;0.5)\% \;,
\end{equation}

\begin{equation}
B(B\rightarrow X\mu\nu ) = (10.3\;\pm \;0.5)\% \;,
\end{equation}

\begin{equation}
B(B\rightarrow X\tau\nu ) = (2.8\;\pm \;0.6)\% \;,
\end{equation}

\begin{equation}
\sum_\ell B(B\rightarrow X\ell\nu ) = (23.8\;\pm \;0.9)\% \;.
\end{equation}
Putting it all together we estimate

\begin{equation}
B(b\rightarrow c\bar cs) \;\approx \;0.31 \;,
\end{equation}
\begin{equation}
B(b\rightarrow c\bar cs') \;\approx \;0.33 \;.
\end{equation}
We confirm the theoretical
expectation~\cite{baffle,fwd,voloshin,bagan} that
\begin{equation}
n_c \;\approx \; 1\;+\;B(b\rightarrow c\bar cs')\;\approx \;1.3 \;,
\end{equation}
and predict
\begin{equation}
(| \Delta\Gamma |/\Gamma )_{B_s}\;\raisebox{-.4ex}
{\rlap{$\sim$}} \raisebox{.4ex}{$<$}\; B(b\rightarrow c\bar cs) \;\approx \;
0.31 \;.
\end{equation}

Strictly speaking, however, it becomes meaningless to speak about branching
fractions of $B^0$ to final states $f$, because one does not know which width
$\Gamma_L$ or $\Gamma_H$ is to be used in the denominator.
The situation is completely analogous to the neutral kaons. We therefore will
have to talk about the branching fractions of the heavy and light $B_s$
mesons to final states $f$, i.e.
$B(B_{H,L}\rightarrow f)$.
For instance, the semileptonic widths satisfy
\begin{eqnarray}
B(B_L \rightarrow D^{(*)-}_s \ell^+\nu ) & = & \frac{\Gamma (B_L \rightarrow
D^{(*)-}_s
\ell^+ \nu )}{\Gamma_L} = \frac{|p|^2 \;\Gamma (B^0 \rightarrow D^{(*)-}_s
\ell^+
\nu )}{\Gamma_L} \approx \nonumber \\
& \approx & \frac{\Gamma (B^0 \rightarrow D^{(*)-}_s \ell^+ \nu )}{2\Gamma_L}
\;,
\end{eqnarray}
\begin{eqnarray}
B(B_H \rightarrow D^{(*)-}_s \ell^+\nu ) & = & \frac{\Gamma (B_H \rightarrow
D^{(*)-}_s
\ell^+ \nu )}{\Gamma_H} = \frac{|p|^2 \;\Gamma (B^0 \rightarrow D^{(*)-}_s
\ell^+
\nu )}{\Gamma_H} \approx \nonumber \\
& \approx & \frac{\Gamma (B^0 \rightarrow D^{(*)-}_s \ell^+ \nu )}{2\Gamma_H}
\;.
\end{eqnarray}
Whereas the numerators are identical, the denominators may differ
substantially which causes different (in our example, semileptonic) branching
fractions of the heavy and light $B_s$.
Further, a sizable width difference allows CP violating measurements and the
clean extraction of CKM-phases with \underline{untagged} $B_s$ data samples,
which will be expanded upon below.  Clearly, the observation of a large width
difference in $B_s$ mesons will have
important ramifications for the Standard Model.  Because establishing a
non-vanishing width difference is so important, this note lists a few
suggestions in how to measure $(\Delta \Gamma)_{B_s}$.  To reach that goal,
Section IV reviews time-dependences of $B^0$ decays.

\section{Time Dependences}

This section gives a set of master equations from which one can read off
desired time-dependences.  Denote by $B^0_{phys}\; (\overline{B}^0_{phys}$)
a time-evolved
initially unmixed $B^0 \; (\overline{B}^0)$.
\begin{equation}
\mid B^0_{phys} (t=0)\rangle = | B^0 \rangle \;.
\end{equation}
Consider final states $f$ which can be fed from both a
$B^0$ and a $\overline{B}^0$,
and define the interference terms

\begin{equation}
\lambda \equiv \frac{q}{p}\;\frac{ \langle f\mid\overline{B}^0
\rangle }{ \langle f\mid B^0 \rangle},\;
\overline{\lambda} \equiv \frac{p}{q} \;\frac{ \langle \overline f \mid
B^0 \rangle}{ \langle \overline f \mid\overline{B}^0  \rangle }.
\end{equation}
Without any assumptions, the time-dependent rates are given
by~\cite{dr,dthesis}
\begin{eqnarray}
\Gamma (B^0_{phys} (t)\rightarrow f ) & = & \Gamma (B^0 \rightarrow f)
\bigg\{\mid g_+ (t)\mid^2 +\mid\lambda\mid^2 \;\mid g_-(t)\mid^2 + \nonumber \\
& + & 2 Re\left[\lambda \;g_-(t) \;g^*_+ (t)\right]\bigg\}\;,
\end{eqnarray}
\begin{eqnarray}
\Gamma (B^0_{phys} (t) \rightarrow  \overline f   ) & = & \Gamma
(\overline{B}^0
\rightarrow  \overline f   ) \;\left|\frac{q}{p}\right|^2\bigg\{\mid g_-
(t)\mid^2 +
\mid\overline{\lambda}\mid^2 \;\mid g_+ (t)\mid^2 + \nonumber \\
& + & 2 Re \left[\overline{\lambda} \;g_+ (t)\; g^*_- (t)\right]\bigg\}\;,
\end{eqnarray}
\begin{eqnarray}
\Gamma (\overline{B}^0_{phys}(t)\rightarrow \overline f   ) & = & \Gamma(
\overline{B}^0 \rightarrow \overline f   ) \;\bigg\{ \mid g_+(t)\mid^2 + \mid
\overline{\lambda}\mid^2 \;\mid g_-(t)\mid^2 + \nonumber \\
& + & 2 Re \left[ \overline{\lambda}\; g_- (t)\; g^*_+ (t) \right]\bigg\}\;,
\end{eqnarray}
\begin{eqnarray}
\Gamma (\overline{B}^0 _{phys} (t)\rightarrow f) & = & \Gamma (B^0 \rightarrow
f)
\;\left|\frac{p}{q}\right|^2 \;\bigg\{ \mid g_-(t)\mid^2 + \mid\lambda
\mid^2 \;\mid g_+ (t)\mid^2 + \nonumber \\
& + & 2 Re \left[\lambda \;g_+(t)\;
 g^*_- (t) \right]\bigg\}\; ,
\end{eqnarray}
where
\begin{equation}
\mid g_\pm (t)\mid^2 = \frac{1}{4} \;\bigg\{ e^{-\Gamma_L t} +e^{-\Gamma_H t}
\pm
2 e^{-\Gamma t} \cos\Delta mt\bigg\} \;,
\end{equation}

\begin{equation}
g_-  (t)\; g^*_+ (t) =\frac{1}{4} \;\bigg\{ e^{- \Gamma_L t} -e^{-\Gamma_H t} +
 2i \;e^{-\Gamma t} \sin\Delta mt\bigg\} \;.
\end{equation}

Those equations make a very important point transparent. For
$\left|\frac{q}{p}\right|=1$, the rapid time-dependent oscillations
dependent on $\Delta mt$ cancel in untagged data samples,

\begin{equation}
\Gamma\left[f\left(t\right)\right]\equiv\Gamma\left(B^0_{phys}\left(t\right)
\rightarrow f\right)+\Gamma\left(\overline{B}^0_{phys}\left(t\right)\rightarrow
f\right) \;,
\end{equation}

\begin{eqnarray}
\Gamma \left[f\left(t\right)\right] & = &  \frac{\Gamma (B^0 \rightarrow
f)}{2} \;\bigg\{\left(1 + |\lambda |^2 \right) \;\left(e^{-\Gamma_L t}
+e^{-\Gamma_H t}\right) + \nonumber \\
& + & 2 Re \;\lambda \;\left(e^{-\Gamma_L t} -e^{-\Gamma_H t}
\right)\bigg\} \;,
\end{eqnarray}

\begin{eqnarray}
\Gamma \left[\overline f\left(t\right)\right] & = &  \frac{\Gamma (\overline
B^0 \rightarrow
\overline f)}{2} \;\bigg\{\left(1 + |\overline \lambda |^2 \right)
\;\left(e^{-\Gamma_L t}
+e^{-\Gamma_H t}\right) + \nonumber \\
& + & 2 Re \;\overline \lambda \;\left(e^{-\Gamma_L t} -e^{-\Gamma_H t}
\right)\bigg\} \;.
\end{eqnarray}
The only time-dependences remaining are that of the two exponential falloffs,
$e^{-\Gamma_{L,H}t}$, both of which are at the average $b$-lifetime scale.
{}From the two time-scales---$1/\Delta m$ and $1/\Gamma$---governing
time-dependent $B_s$ decays, choosing untagged data samples removes any
dependence on the much shorter $1/\Delta m$ scale,
$$1/\Delta m \ll 1/\Gamma \;\; .$$
This is of prime importance on several counts.  First, at $e^+e^-$ and $p\bar
p$
colliders any $B_s$ candidate belongs automatically to the untagged data
sample. Tagging this event will cost in efficiency and in purity. Collecting an
untagged data sample at $pp$ colliders or fixed target experiments can be done
but is more involved and will not be addressed here.
Second, $\Delta m/\Gamma$ could turn out to be larger than what present
technology can resolve, although there exists an intriguing expression of
interest for a forward collider experiment \cite{butler} that claims to be able
to
study $\Delta m/\Gamma \;\raisebox{-.4ex}{\rlap{$\sim$}} \raisebox{.4ex}{$<$}\;
60$
which is above the upper CKM-model limit \cite{xsnir}.

We wish to present some theorems which will be used throughout this note.  For
that
purpose, define
\begin{equation}
\mid  \overline f  \rangle \equiv CP\mid f \rangle  , \mid\overline{B}^0
 \rangle \;\equiv CP\mid B^0\rangle \;.
\end{equation}
Suppose that a unique CKM combination governs $B^0 \rightarrow f$ and
another unique one $\overline{B}^0 \rightarrow f$, then the following
Theorems and consequences hold.

\begin{flushleft}
\underline{Theorem 1}
\end{flushleft}

If the amplitude for $B^0 \rightarrow f$ is denoted by
\begin{equation}
 \langle f \mid B^0\rangle =G\mid a\mid e^{i\delta} \;,
\end{equation}
then the CP-conjugated amplitude is
\begin{equation}
 \langle  \overline f   \mid\overline{B}^0  \rangle  = G^* \mid a\mid
e^{i\delta} \;.
\end{equation}
Here $G$ is the unique CKM-combination, $\mid a\mid$ the magnitude of the
strong
matrix element, and $\delta$ a possible strong interaction phase.

\begin{flushleft}
\underline{Consequence 2}
\end{flushleft}

\begin{equation}
| \langle f|B^0 \rangle | = | \langle  \overline f   |\overline{B}^0 \rangle|
\;.
\end{equation}

\begin{flushleft}
\underline{Consequence 3}
\end{flushleft}

If furthermore $\left|\frac{q}{p}\right| \approx 1$ is assumed, then
\begin{equation}
\lambda =|\lambda | \;e^{i(\phi +\Delta )} \; ,
\end{equation}
\begin{equation}
\overline{\lambda} =|\lambda | \;e^{i(-\phi +\Delta )} \;.
\end{equation}
where $\phi$ denotes the CKM phase, and $\Delta$ the possible
strong interaction phase difference.

\begin{flushleft}
\underline{Consequence 4}
\end{flushleft}

Consider final states $f$ which are CP eigenstates governed by the same unique
CKM combination. The sign of the
interference term flips, depending on the CP-parity of $f$,
\begin{equation}
\lambda_{CP=+} =-\lambda_{CP=-} \;.
\end{equation}

\begin{flushleft}
\underline{Theorem 5}
\end{flushleft}

If in addition $\left|\frac{q}{p}\right| =1$ is assumed,
then for a CP-eigenstate $f$ (either CP-even or CP-odd),
\begin{equation}
\overline{\lambda} =\lambda^* , \;\;\text{and} \;\;|\lambda | =1 \;.
\end{equation}

Although the proofs of the theorems and consequences are well
known~\cite{cpreviewBigi}, they will be rederived here for completeness sake
and to illuminate what is exactly meant by final state phase differences.
The proof of Theorem 1 is based on the fact that CP violation occurs
only due to complex-valued CKM elements within the CKM model. The
Hamiltonian which governs $B^0\rightarrow f$ decays can thus be
factorized as,
\begin{equation}
{\cal H} = Gh +G^* h^+ \;.
\end{equation}
Here $h$ is the sum of all relevant operators annihilating a $B^0$
and creating $f$, schematically written as (for example)
\begin{equation}
h=(\overline{b}c)_{V-A} \;\;(\overline{u}s)_{V-A} \;.
\end{equation}
The hermitian conjugate $h^+$ annihilates a $\overline{B}^0$ and creates
$\overline f$.
Since CP-violation resides solely within the CKM elements,
the $h$'s satisfy
\begin{equation}
(CP)^+ \;h\;CP = h^+ , \;\;(CP)^+ \; h^+ \;CP = h \;.
\end{equation}
Now, the amplitude of $B^0$ to $f$ stands actually for
\begin{equation}
\langle f|B^0 \rangle \equiv \langle f|{\cal H}|B^0\rangle = G\langle
f|h|B^0 \rangle
=G|a|e^{i\delta} \;.
\end{equation}
The strong matrix element is
\begin{equation}
\langle f|h|B^0\rangle = |a|e^{i\delta} \;.
\end{equation}
The CP-conjugated amplitude satisfies (using Eqs.~(4.20), (4.12), (4.22),
(4.24) in the second, third,
fourth, and fifth step, respectively),
\begin{eqnarray}
\langle \overline f |\overline{B}^0 \rangle & \equiv & \langle\overline f
|{\cal H}|
\overline{B}^0 \rangle = G^*
\langle \overline f |h^+ |\overline{B}^0 \rangle = \nonumber \\
& = & G^* \langle f|(CP)^+ h^+ CP|B^0 \rangle = G^* \langle f|h|B^0 \rangle =
\nonumber \\
& = & G^* |a| e^{i\delta} \;.
\end{eqnarray}
Theorem 1 is thus proven, and Consequence 2 results immediately. Consequence 3
is proven as follows. Denote the amplitude of $B^0\rightarrow f$ as
\begin{equation}
\langle f|B^0 \rangle = G|a|e^{i\delta} \;,
\end{equation}
and that of $B^0 \rightarrow \overline f$ as
\begin{equation}
\langle \overline f |B^0 \rangle = K |b|e^{i\tau} \;,
\end{equation}
where $G, K$ are the unique CKM-combinations, $|a|, |b|$ magnitudes of strong
matrix
elements, and $\delta ,\tau$ their respective strong phases. Theorem
1 informs us that
\begin{equation}
\langle \overline f |\overline B^0 \rangle = G^*|a|e^{i\delta} \;,
\end{equation}
\begin{equation}
\langle f|\overline{B}^0 \rangle = K^* |b |e^{i\tau} \;.
\end{equation}
{}From the definitions of the interference terms,
\begin{equation}
\lambda \equiv \frac{q}{p} \;\frac{\langle f|\overline B^0 \rangle}{\langle
f|B^0 \rangle} = \frac{q}{p} \;\frac{K^*}{G} \;\frac{|b|}{|a|}
\;e^{i(\tau -\delta)}\;,
\end{equation}
\begin{equation}
\overline \lambda \equiv \frac{p}{q} \;\frac{\langle \overline f |B^0 \rangle}
{\langle \overline f |\overline B^0 \rangle} = \frac{p}{q} \; \frac{K}{G^*} \;
\frac{|b|}{|a|} \;e^{i(\tau -\delta )} \;.
\end{equation}
Because $\left|\frac{q}{p}\right| =1,$ we get $p/q = (q/p)^*$ and
\begin{equation}
\lambda = \lambda_{CKM} \;z \;, \;\;\overline \lambda = \lambda^*_{CKM} \;z \;.
\end{equation}
The CKM combination of the interference term is denoted by
\begin{equation}
\lambda_{CKM} = \frac{q}{p} \;\frac{K^*}{G} \;\equiv \;|\lambda_{CKM}|
\;e^{i\phi} \;,
\end{equation}
whereas the ratio of strong matrix elements is
\begin{equation}
z \;\equiv \;\left|\frac{b}{a}\right| \;e^{i(\tau -\delta )}
\;\equiv \;|z| \;e^{i\Delta} \;.
\end{equation}
Consequence 3 is proven, where
$\Delta \equiv \tau -\delta$ denotes
the phase difference between the two strong matrix elements. To prove
Consequence 4, consider a CP-eigenstate  $f_\eta$ with CP-parity
$\eta \;(=\pm\; 1)$. As before, define
\begin{equation}
\langle f_\eta |B^0 \rangle = G|a| \;e^{i\delta} \;.
\end{equation}
Theorem 1 yields
\begin{equation}
\eta \langle f_\eta |\overline B^0 \rangle = G^* |a| \;e^{i\delta} \;,
\end{equation}
and
\begin{equation}
\lambda_\eta =\frac{q}{p} \;\frac{\langle f_\eta |\overline B^0
\rangle}{\langle
f_{\eta} |B^0\rangle} =\eta \;\frac{q}{p} \;\frac{G^*}{G} \;.
\end{equation}
That is,
\begin{equation}
\lambda_+ =\frac{q}{p} \;\frac{G^*}{G} \;, \;\;\lambda_- =-\lambda_+
\end{equation}
and Consequence 4 is proven.
Proving Theorem 5 is also straightforward.  We get
\begin{equation}
\langle \overline{f}_\eta |B^0 \rangle =\eta \;\langle f_\eta |B^0 \rangle \;,
\end{equation}
\begin{equation}
\langle \overline{f}_\eta |\overline{B}^0 \rangle = \eta \;\langle f_\eta |
\overline{B}^0 \rangle \;.
\end{equation}
Since
\begin{equation}
\lambda \equiv \;\frac{q}{p} \;\frac{\langle f_\eta |
\overline{B}^0 \rangle}{\langle f_\eta |B^0 \rangle} \;, \;\;\text{and}
\end{equation}
\begin{equation}
\overline \lambda \equiv \;\frac{p}{q} \;\frac{\langle \overline f_\eta
|B^0 \rangle}{\langle \overline f_\eta |\overline B^0 \rangle}
= \frac{p}{q} \;\frac{\langle f_\eta |B^0 \rangle}{\langle f_\eta
| \overline B^0 \rangle} = \frac{1}{\lambda} = \lambda^* \;.
\end{equation}
The second and third steps in Eq.~(4.42) occur because of Eqs.~(4.39)-(4.40)
and (4.41)
respectively. The last step occurs because $|\lambda |^2 =1$, which happens
since $\left|\frac{q}{p}\right|=1$ is assumed and $\left|\frac{\langle f_\eta
|\overline B^0 \rangle}{\langle f_\eta |B^0 \rangle}\right| =1$ due to
Eqs.~(4.35)-(4.36) or equivalently due to Consequence 2.

Consider the situation under which the above-mentioned theorems
and consequences hold (i.e., a unique CKM combination governs
$B^0\rightarrow f$ and another unique one $\overline{B}^0 \rightarrow f$) and
assume $\left|\frac{q}{p}\right| =1$, then the time-dependent rates
simplify from Eqs.~(4.3) - (4.6) to:
\begin{equation}
\Gamma \left(B^0_{phys}\left(t\right)\rightarrow f\right) =\Gamma \left(
B^0 \rightarrow f\right) \bigg\{|g_+ \left(t\right)|^2 +|\lambda |^2\;
|g_- \left(t\right)|^2 + 2Re \left[\lambda\; g_-\left(t\right)
g^*_+\left(t\right)
\right]\bigg\} \;,
\end{equation}
\begin{equation}
\Gamma \left(B^0_{phys}\left(t\right)\rightarrow  \overline f  \right) =\Gamma
\left(B^0 \rightarrow
f\right)\bigg\{|g_-\left(t\right)|^2 +|\lambda |^2 \;|g_+ \left(t\right)|^2
+2Re\left[\overline{\lambda}
\;g_+\left(t\right)g_-^*\left(t\right)\right]\bigg\}\;,
\end{equation}
\begin{equation}
\Gamma\left(\overline{B}^0_{phys}\left(t\right)\rightarrow  \overline f
\right)=\Gamma
\left(B^0 \rightarrow f\right)\bigg\{\left|g_+\left(t\right)\right|^2
+|\lambda |^2 \;|g_-
 \left(t\right) |^2 + 2Re\left[\overline{\lambda}\;g_-\left(t\right) g^*_+
\left(t\right)\right]\bigg\}\;,
\end{equation}
\begin{equation}
\Gamma\left(\overline{B}^0_{phys} \left(t\right)\rightarrow f\right)=\Gamma
\left(B^0 \rightarrow f\right)\bigg\{|g_-\left(t\right)|^2 +|\lambda |^2 \;|g_+
\left(t\right)|^2 +2Re\left[\lambda
\;g_+\left(t\right)g^*_-\left(t\right)\right]
\bigg\} \;.
\end{equation}
The above four equations are our master equations. By considering
different cases, the next section demonstrates how untagged data samples
of $B_s$ mesons could be used not only to extract the light and heavy widths,
but even
the unitarity angle $\gamma$ and CP-violation.

\section{Physics with Modes of Untagged $B_s$ Mesons}

Unless explicitly stated otherwise,
this section supposes that the conditions hold under which the master
equations,
Eqs.~(4.43) - (4.46), are satisfied---that is, $\left|\frac{q}{p}\right| =1$
and unique CKM combinations govern the decays of the unmixed $B_s$
and $\overline{B_s}$
to $f$.
We analyze the time-dependences for several cases of untagged $B_s$ data
samples.  First, flavor-specific modes $g$ of $B_s$ are studied, such that an
unmixed
$B_s$ decays to $g$, whereas an unmixed
$\overline{B_s}$ is never seen in $g$,
$\overline B_s \not\rightarrow g$.  Examples for $g$ are
$D_s^{(*)-} \ell^+\nu ,D^{(*)-}_s \pi^+ , D^{(*)-}_s a^+_1 , D^{(*)-}_s
\rho^+$.

Second, time-evolutions of CP eigenmodes of $B_s$ mesons are scrutinized.
Within the CKM model, CP-eigenmodes of $B_s$ decays driven by $b\rightarrow
c\bar c s$ are governed by a single exponential decay law.  In contrast, there
are CP-eigenmodes that are governed by two exponential decay laws, which
signals CP violation.  A time-dependent study of the untagged $\rho^0 K_S$ data
sample extracts the angle $\gamma$ of the CKM unitarity triangle, when penguin
amplitudes can be neglected.  The penguin amplitudes are in general
non-negligible for $B_s \to \; \rho^0 K_S$.

We discuss thus next the extraction of $\gamma$ from modes $f$ that can be fed
from both $B^0_s$ and $\overline{B^0_s}$, such as
$D^{(*)-}_s K^{(*)+}, \overline{D}^{(*)0}\phi , \overline{D}^{(*)0}\eta$.
Sizable CP violating effects could be seen when untagged time-evolutions of $f$
are compared with those of $\overline f$.  We then investigate what occurs when
several CKM-combinations contribute to the decay-amplitude of an unmixed $B_s$.
The last subsection combines all the information and spells out many methods
for measuring a width difference from untagged $B_s$ samples.  Some of the
methods are directly applicable to the current flavor-specific world data
sample of $B_s$.

\subsection{Flavor-Specific Modes of $B_s$}

Since only the unmixed $B^0$ can be seen in $g$, but never the unmixed
$\overline{B}^0 $, one obtains
\begin{equation}
\lambda =\overline{\lambda} =0 \;.
\end{equation}
The time-dependent rates become \cite{dr,shifman,dpt,dthesis}
\begin{equation}
\Gamma \left(B^0_{phys}\left(t\right)\rightarrow g\right) = \Gamma
\left(B^0 \rightarrow g\right) \left|g_+\left(t\right)\right|^2 \;,
\end{equation}
\begin{equation}
\Gamma \left(\overline{B}^0_{phys}\left(t\right)\rightarrow g\right) =
\Gamma\left(B^0 \rightarrow g\right) \left|g_-\left(t\right)\right|^2 \;,
\end{equation}
and
\begin{equation}
\Gamma\left[\overline g\left(t\right)\right] =
\Gamma\left[g\left(t\right)\right]=\frac{\Gamma\left(B^0 \rightarrow
g\right)}{2} \;
\bigg\{e^{-\Gamma_L t} +e^{-\Gamma_H t} \bigg\} \;.
\end{equation}
The untagged time-dependent rates for the process and CP-conjugated process are
the same.
The untagged data sample time-evolves as the sum of two exponentials
\cite{dpt,qpnot1}.
Examples for such flavor specific modes $g$ are
\begin{equation}
D_s^{(*)-} \ell^+\nu , \;\;D_s^{(*)-} \pi^+ , \;\; D_s^{(*)-} a_1^+ ,
\;\; D_s^{(*)-} \rho^+ \;.
\end{equation}
More than 200 such $B_s$-events have been recorded at CDF \cite{bscdf}
and the LEP \cite{bslep} experiments.  Their time-dependence has been fit
to a single exponential, which essentially measures the average $B_s$ width
$\Gamma$~\cite{feasibility}. This measurement for $\Gamma$ could then be used
to determine $\Delta\Gamma$ by fitting the time-evolution of the same data
sample to the correct functional form,
\begin{equation}
e^{-\left(\Gamma + \frac{\Delta\Gamma}{2}\right)t} + e^{-\left(\Gamma -
\frac{\Delta\Gamma}{2}\right)t} \;.
\end{equation}

\subsection{CP Eigenstates}

This subsection considers modes $f$ of $B_s$ that have definite CP parity.
The CP-even (CP-odd) final state will sometimes be denoted as
$f_+ \;(f_- )$. We first describe how to determine $\Gamma_L$ from the
CP-even modes governed
by the $b\rightarrow c\bar c s$ transition.
The CP-odd modes driven by $b\rightarrow c\bar c s$ are governed by the
$e^{-\Gamma_H t}$ exponent, and allow the determination of $\Gamma_H$,
in principle.  The CP-odd modes however are not only predicted to be rarer
than the CP-even modes, but are harder to detect.  One possible determination
of $\Delta\Gamma$ could use the
largest $B_s$ data sample, that of flavor specific decays of $B_s$, combined
with the above-mentioned measurement of $\Gamma_L$ to extract $\Gamma_H$.
The CP-odd modes driven by $b\rightarrow c\bar c s$ are governed by the
exponent $\exp(-\Gamma_H t)$ and may be used as a consistency check to
determine $\Gamma_H$.
Once a width difference between $\Gamma_H$ and $\Gamma_L$ has been
established, interesting CP violating effects and the clean extraction
of fundamental CKM-parameters become possible with untagged $B_s$ data samples.

CP invariance requires a single exponential decay law for tagged and
untagged neutral $B$'s seen in a CP eigenstate.
The CKM model predicts two different exponential decay laws for many CP
eigenstates of $B_s$ decays, such as
$\rho^0 K_S , \;D^0_{CP} \phi ,\;K^+ K^-$.
Not only can CP violation be exhibited, but even CKM-phases can be extracted
from time-dependent studies of \underline{untagged} $B_s$ data samples.
For instance, the time-evolution of the \underline{untagged} $\rho^0 K_S$ mode
extracts $\cos (2\gamma )$ as shown below, when penguin contributions are
neglected.
Penguins may be sizable however, in which case one may use non-CP
eigenmodes to extract $\gamma$ as will be discussed in the next subsection.

Suppose that a unique CKM-combination governs the decay of $B^0$ to
CP-eigenstate
$f$ and that $\left|\frac{q}{p}\right|=1$, then the time-dependent rates
become:
\begin{eqnarray}
\Gamma\left(B^0_{phys} \left(t\right)\rightarrow f\right) & = &
\frac{\Gamma (B^0 \rightarrow f )}{2} \;\bigg\{e^{-\Gamma_L t} +e^{-\Gamma_H t}
+ \nonumber \\
& + & Re\lambda \left(e^{-\Gamma_L t} -e^{-\Gamma_H t}\right)
 -2 \;Im\lambda\; e^{-\Gamma t}
\sin\Delta mt\bigg\} \;,
\end{eqnarray}
\begin{eqnarray}
\Gamma\left(\overline{B}^0_{phys}\left(t\right)\rightarrow f\right) & = &
\frac{\Gamma (B^0 \rightarrow f)}{2}\;\bigg\{ e^{-\Gamma_L t}
+e^{-\Gamma_H t} + \nonumber \\
& + & Re\lambda\left(e^{-\Gamma_L t}-e^{-\Gamma_H t}\right) +2 Im\lambda \;
e^{-\Gamma t} \sin\Delta mt\bigg\} \;.
\end{eqnarray}
As advertised, the rapid $\Delta mt$ oscillations cancel in the time-dependent
rate of the untagged data sample,
\begin{equation}
\Gamma\left[f\left(t\right)\right]  =\Gamma\left(B^0 \rightarrow
f\right)\bigg\{
e^{-\Gamma_L t} + e^{-\Gamma_H t} +
Re\lambda \left(e^{-\Gamma_L t} -e^{-\Gamma_H t}\right)\bigg\} \;.
\end{equation}
CP-violating effects are predicted to be small for CP-eigenmodes of $B_s$
governed by $b\rightarrow c\bar c s$ \cite{dr,ddgn,dsnowmass},
\begin{equation}
0.01 \raisebox{-.4ex}{\rlap{$\sim$}} \raisebox{.4ex}{$<$}\; Im\lambda \;
\raisebox{-.4ex}{\rlap{$\sim$}} \raisebox{.4ex}{$<$}\; 0.05 \;.
\end{equation}
Since here $|\lambda |\approx 1$ to excellent accuracy, we obtain
\begin{equation}
0.999 \;\raisebox{-.4ex}{\rlap{$\sim$}} \raisebox{.4ex}{$<$}\;|Re\lambda |
< 1 \;.
\end{equation}

Eq.~(5.11) tells us that the untagged data sample of CP-eigenmodes
of $B_s$ governed
by $b\rightarrow c\bar c s$ involves unobservably tiny CP violating effects.
In the absence of CP-violation, the CP-even (CP-odd) interference term is
\begin{equation}
\lambda_+ =1 \;\;(\lambda_- =-1)\;.
\end{equation}
The time-dependence of the untagged data sample is
\begin{equation}
\Gamma \left[f_+ \left(t\right)\right] =2\Gamma \left(B^0 \rightarrow f_+
\right) \;e^{-\Gamma_L t} \;,
\end{equation}
\begin{equation}
\Gamma \left[f_- \left(t\right)\right] =2\Gamma \left(B^0 \rightarrow f_-
\right) \;e^{-\Gamma_H t} \;,
\end{equation}
and the CP-even rate is identified with $\Gamma_L$,
\begin{equation}
\Gamma_+ =\Gamma_L
\end{equation}
and the CP-odd rate with $\Gamma_H$,
\begin{equation}
\Gamma_- =\Gamma_H
\end{equation}
This is consistent with the assignment made in Eq.~(2.17). Aleksan et al.
\cite{aleksan}
claimed to have shown that $\Gamma_+ -\Gamma_-> 0$ from both a box
diagram calculation and from a sum over many exclusive modes. Our
addition, in that respect, is the identification $\Gamma_H =\Gamma_-$
and $\Gamma_L =\Gamma_+$.
Examples of modes with even CP-parity are
$J/\psi\eta ,\;D^+_s D^-_s$.
It is not easy to come up with CP-odd modes, for example
$J/\psi f_0 (980) , \;J/\psi a_0 (980)$.
 In contrast, $J/\psi\phi ,\;D^{*+}_s D^{*-}_s ,\;
D^{*+}_s D^{-}_s + D^+_s D^{*-}_s$
are dominantly CP-even \cite{rosnerdecayconstant,aleksan}, with possibly small
CP-odd components.
The evidence that the $J/\psi \phi$ mode is mainly CP-even comes from the
observed angular correlations of the $B\rightarrow J/\psi K^*$ mode
\cite{psikstar} coupled with $SU(3)$ flavor symmetry~\cite{dsnowmass}, or from
an
explicit calculation assuming factorization \cite{aleksan}.
In any event, an angular correlation study separates in general
the CP-even and CP-odd components \cite{kkps,dqstl}. Once the CP-even
and CP-odd components have been separated, their different lifetimes
could be determined \cite{bigirosner}. In practice, however, the CP-odd modes
occur much
less frequently than the CP-even modes, and are harder to detect.
Thus, $\Gamma_L$ will be known well, whereas $\Gamma_H$ could be obtained
from the time-evolution of untagged, flavor-specific modes $g$ of $B_s$,
\begin{equation}
\Gamma \left[ \overline g \left(t\right)\right] +
\Gamma\left[g\left(t\right)\right]\sim e^{-\Gamma_H t} +e^{-\Gamma_L t} \;.
\end{equation}
Examples of $g$ have been listed in the previous subsection, in which
$D_s$ is dominantly featured. A discriminating feature between $D_s$ and other
charmed hadrons is the inclusive $\phi$ yield.
Whereas the inclusive $\phi$ yield in $D_s$ decays is about 20\% or more,
it is much smaller in $D^+$ and $D^0$ decays \cite{pdg,distinguish}.
Mainly due to this large inclusive $\phi$ yield in $D_s$ decays, and partly
because $\phi$ even appears in the $B_s \rightarrow J/\psi\phi$ mode, we
strongly support the use of a $\phi$ trigger in experimental
studies~\cite{phitrigger}.

Although $\rho^0 K_S$ is CP-odd, it is
in general not governed by a single exponential decay law, because its
interference term satisfies~\cite{dr,ddw}
\begin{equation}
Re\lambda =-\cos (2\gamma ) \;,
\end{equation}
when penguin amplitudes are neglected.
Time-dependences of \underline{untagged} $\rho^0 K_S$ events extract
$\cos (2\gamma )$; see Eq.~(5.9).  They exhibit CP violation when more than one
exponential decay law contributes.  Far reaching consequences on the CKM-model
would result, even if the  $\rho^0 K_S$ mode were governed by a single
exponential decay law.  The interference term would satisfy $Re\lambda = \pm
1$.  If $Re\lambda = + 1$, then the CP-odd $\rho^0 K_S$ decay mode is governed
by $\Gamma_L$.  This constitutes a clear violation of CP, because the
time-evolution of the CP-odd mode $\rho^0 K_S$ is governed by the same exponent
$\Gamma_L$ as the opposite CP-even modes driven by $b\rightarrow c\bar c s$
(and not by $\Gamma_H$ governing CP-odd modes driven by $b\rightarrow c\bar c
s$).  On the other hand, if $Re\lambda = - 1$, then $\sin\gamma = 0$,
contradicting what is currently known about $\sin\gamma$ in the
CKM-model,~\cite{stone}
\begin{equation}
0.5 \;\raisebox{-.4ex}{\rlap{$\sim$}} \raisebox{.4ex}{$<$}\; \sin\gamma \leq 1
\;.
\end{equation}

Penguin amplitudes may be significant however, in which case several
CKM-combinations contribute to the unmixed amplitude.  The time-dependent,
untagged  decay-rate
(assuming $\left|\frac{q}{p}\right| =1$) becomes
\begin{eqnarray}
\Gamma\left[f\left(t\right)\right] & = & \Gamma\left(B^0 \rightarrow
f\right)\bigg\{\frac{1}{2}\left(e^{-\Gamma_L t} +e^{-\Gamma_H t}\right)
\left(1+|\lambda |^2\right) + \nonumber \\
& + & Re\lambda\left(e^{-\Gamma_L t} -e^{-\Gamma_H t}\right)\bigg\} \;.
\end{eqnarray}
This equation is relevant to, for instance, the $\rho^0 K_S$, $D^0_{CP}\phi,
\;K^+K^- , \phi K_S$ modes of $B_s$. It shows that those CP-eigenmodes will
have in general two
exponential decay laws, which demonstrates CP violation. Other relevant,
experimentally
accessible modes are $\phi\phi,\; \rho^0\phi$. Angular correlations can
separate their
CP-even and CP-odd components \cite{kkps,dqstl}. If any component with definite
CP-parity has two exponential decay laws, CP-violation occurs.  CP violation
may be seen not only in definite CP-components, but in interference effects
between different helicity amplitudes.

Because of a possible penguin contamination, the unitarity
angle $\gamma$ cannot be extracted cleanly from the time-evolution of
untagged $\rho^0 K_S$ events. In contrast, a clean extraction is possible
from non-CP eigenmodes which do not suffer from penguin contamination
at all, as discussed next.

\subsection{Modes Common to $B_s$ and $\overline{B_s}$}

It is well known \cite{glDphi,adk,Dsk,aleksansumDsk} that tagged,
time-dependent studies (capable of observing the rapid $\Delta
mt$-oscillations) are able to extract the unitarity angle $\gamma$ and observe
CP violation from $B_s$-modes governed by the $b\rightarrow c\bar u s, u \bar c
s$ transitions, such as
$$f=D^{(*)-}_s K^{(*)+}, \;\overline{D}^{(*)0}\phi , \;\overline{D}^{(*)0}\eta
\;.$$
This subsection demonstrates that even \underline{untagged}, time-dependent
studies (now governed only by the two exponential decay laws) are able to
extract the angle $\gamma$.  Those untagged studies may observe CP violation
for non-vanishing strong final-state phase differences.  A non-zero strong
final-state phase difference could arise from traditional rescattering effects
or from resonance effects discovered recently by Atwood et al. in a different
context~\cite{atwoodsoni,atwood}.  For traditional rescattering effects, CP
violation
 is probably more pronounced in color-suppressed modes,
$\overline{D}^{(*)0} \phi ,\;\overline{D}^{(*)0} \eta$, than in the
color-allowed
ones, $D^{(*)-}_s K^{(*)+}$.
The reason is simple.  Within the factorization approximation
\cite{bsw,bijnens},
rates of color-suppressed modes are tiny with respect to color-allowed ones.
The latter may rescatter into the former causing possibly sizable strong
phase differences $\Delta$ for the color-suppressed modes.
In contrast, such large rescattering effects are not likely to occur for the
color-allowed modes. It is reasonable to expect $\Delta\approx 0$ for
the color-allowed modes.

In a nice series of papers, Atwood et al. have shown how CP violation can be
enhanced by considering modes where several kaon or unflavored resonances
contribute to the final state~\cite{atwoodsoni,atwood}.  ``Calculable"
final-state phases are generated due to the different widths of the resonances.
 A straightforward application of this idea to untagged $B_s$ modes such as
$D_s^{(*)\mp} (K^{(*)} \pi)^\pm, \; D_s^{(*)\mp} (K\rho)^\pm$, enhances CP
violation. Such ``calculable" final-state phases ensure non-vanishing CP
violating effects for the $B_s$-modes of interest here, which are governed by
the
$\overline b\rightarrow \overline c u \overline s$ transition.  The
\underline{untagged} $B_s$ modes, such as $D_s^{(*)\mp} (K^{(*)} \pi)^\pm, \;
D_s^{(*)\mp} (K\rho)^\pm$, may be used to extract the CKM-unitarity angle
$\gamma$.

This subsection is divided into several parts.  First, the angle
$\gamma$ is extracted from time-dependences of \underline{untagged} $B_s$ data
samples such as $D^{(*)\pm}_s K^{(*)\mp}$.
The overall normalization is obtained by assuming factorization for the
color-allowed processes $B_s\rightarrow D^{(*)-}_s K^{(*)+},
D^{(*)-}_s \pi^{(*)+} ,$ where $\pi^{*+}$ denotes $\rho^+, a_1^+ ,$ etc.
One may object to the factorization assumption. We thus determine $\gamma$
from time-dependences of untagged $D^0 \phi , \overline{D}^0 \phi$,
and $D^0_{CP} \phi$ modes.
The determination does not involve any assumption beyond the validity of the
CKM-model. CP violating effects are described next.
By waiting long enough, essentially only the longer lived $B_H$
survives,
$$|B_H \rangle =p|B_s \rangle  - q|\overline B_s \rangle .$$
If the amplitudes $B_s\rightarrow f$ and $\overline B_s \rightarrow f$
are governed by different CKM phases, CP violation may occur.
The relative CKM-phase for $B_s$ modes governed by $\overline b\rightarrow
\bar c u\bar s$ is $\gamma$ and is significant.
Large CP violating effects can be generated, either from
traditional rescattering effects or from resonance effects.

\subsubsection{CKM-phase $\gamma$ from $B_s$ modes governed by
$\overline b \rightarrow \bar c u\bar s$}

The time-evolutions of the untagged $\stackrel{(-)}{f}$ data samples are:
\begin{eqnarray}
\Gamma\left[\stackrel{(-)}{f}\left(t\right)\right] & = & \frac{\Gamma
(B^0\rightarrow f)}{2}\;\bigg\{\left(e^{-\Gamma_L t} +e^{-\Gamma_H t}\right)
\left(1+\left|\lambda\right|^2\right) + \nonumber \\
& + & 2Re\stackrel{(-)}{\lambda}\left(e^{-\Gamma_L t} -e^{-\Gamma_H t}\right)
\bigg\} \;.
\end{eqnarray}
The rapidly oscillating terms of $\Delta mt$ cancel again. A time-dependent
fit extracts
\begin{equation}
\Gamma (B^0 \rightarrow f)\;(1+|\lambda |^2 ), \;\;\Gamma (B^0\rightarrow f)
Re\lambda ,\;\;\Gamma (B^0\rightarrow f)Re\overline{\lambda} \;.
\end{equation}
The overall normalization could be established from the flavor-specific
data sample; see Eq.~(5.4):
\begin{equation}
\Gamma\left[g\left(t\right)\right] =\frac{\Gamma (B^0 \rightarrow
g)}{2} \;\bigg\{e^{-\Gamma_L t} +e^{-\Gamma_H t} \bigg\} \;.
\end{equation}

The ratio of the unmixed rates is well known from theory:
\begin{equation}
\frac{\Gamma (B^0_s\rightarrow D^-_s K^+ )}{\Gamma (B^0_s \rightarrow D^-_s
\pi^+ )}
\approx \left|\frac{V_{us}}{V_{ud}}\right|^2 \;\;\left(\frac{f_K}{f_\pi
}\right)^2
\;\;\text{(phase-space)} \;.
\end{equation}
Here the factorization approximation is used for those color-allowed
modes.  The $W$-exchange amplitude contributing to $B^0_s \rightarrow
D^-_s K^+$ has been neglected \cite{Wexchange} and has been estimated to be
tiny~\cite{aleksan}. It contributes the same unique
CKM-combination as the spectator graph~\cite{adk}.  Future precision studies
would allow incorporation of even those effects.  Analogously, other
theoretically well known
ratios are, for instance,
\begin{equation}
\frac{\Gamma (B^0_s \rightarrow D^{(*)-}_s K^{(*)+} )}
{\Gamma (B^0_s \rightarrow D_s^{(*)-} \pi^{(*)+} )} \; .
\end{equation}
Combining those well known ratios with the observables in Eq.~(5.22) and the
measured $\Gamma (B^0 \rightarrow g)$ in Eq.~(5.23) extracts:
\begin{equation}
1+|\lambda |^2 \;\;\text{(that is,} \;\;|\lambda |) \;,
\end{equation}
\begin{equation}
Re\lambda =|\lambda | \;\;\cos (\phi +\Delta ) \;,
\end{equation}
and
\begin{equation}
Re\overline{\lambda} =|\lambda | \;\;\cos (-\phi +\Delta )\;.
\end{equation}
Here $\phi =-\gamma$ is the CKM-phase of the interference term $\lambda$
where $\gamma$ is the CKM unitarity angle, and $\Delta$ the strong final
state phase difference.  Finally, the phases $\phi$ and $\Delta$
can be determined up to a discrete ambiguity from $\cos (\phi +\Delta )$ and
$\cos (-\phi +\Delta )$. This implies the determination of the
CKM unitarity-angle $\gamma$ is possible from untagged data samples.
More systematics may cancel by using the ratio
\begin{equation}
\frac{\Gamma\left[\stackrel{\textstyle (-)}{f} \left(t\right)\right]}
{\Gamma\left[ g\left(t\right)\right]} =
\frac{\Gamma (B^0\rightarrow f)}{\Gamma (B^0 \rightarrow g)} \;
\left\{ 1+ |\lambda |^2 +2 Re \stackrel{\textstyle (-)}{\lambda}
\; \tanh \left(\frac{\Delta\Gamma\; t}{2}\right)\right\} \;.
\end{equation}
Theory provides the unmixed ratio $\Gamma (B^0 \rightarrow f)/\Gamma
(B^0 \rightarrow g )$.
The time-independent term yields $|\lambda |$, whereas the time-dependent one
gives $Re\lambda$ and $Re \overline \lambda$. Thus $\phi$ and $\Delta$
can be extracted.

A comment about the discrete ambiguity is in order.  Two solutions
for $\sin^2 \phi$ exist,
\begin{equation}
\sin^2 \phi =\frac{1-c \bar c \pm\sqrt{1+ (c\bar c )^2 -\bar c^2 -c^2}}{2}
\end{equation}
where the extracted cosines are denoted by
\begin{equation}
c=\cos (\phi +\Delta ) \;, \;\;\bar c =\cos (-\phi +\Delta ) \;.
\end{equation}
One solution is the true $\sin^2 \phi$ and the other is the true
$\sin^2 \Delta$.
The CKM-model predicts only large, positive $\sin (-\phi
)=\sin\gamma$~\cite{stone}.
Thus the two-fold ambiguity in $\sin^2\phi$ stays a two-fold ambiguity in
$\sin\phi$, since $\sin \phi \; < \; 0$. Further, this two-fold ambiguity can
be
easily resolved in several ways. First, various final states of $B_s$ driven
by the $\bar b \rightarrow \bar c u\bar s$ transition are governed by
the universal CKM-phase $\phi =-\gamma$. In contrast, they probably will
differ in their strong phase difference $\Delta$. Thus, by considering many
such $B_s$-modes, one can disentangle the universal from the
non-universal phases.  Second, if it were to happen that
$\Delta \approx 0$ for all the many modes, then one solution for
$\sin^2 \phi$ would vanish contradicting Eq.~(5.19). Only one solution for
$\sin^2 \phi$ would remain. This fact can be used to quantify the number of
events required in a feasibility study.  A third way uses resonance effects and
is briefly mentioned below.

For the color-allowed modes, we believe that $\Delta\approx 0$, whereas
for the color-suppressed modes, larger $\Delta$'s could occur.
Thus, $\gamma$ is probably more straightforwardly extracted from the
color-allowed processes, because
\begin{equation}
\cos (\pm\gamma +\Delta )\approx\cos\gamma \;,
\end{equation}
and there may be no need to disentangle $\gamma$ from $\Delta$.

\subsubsection{CKM-phase $\gamma$ from $\overline D^0 \phi, D^0 \phi,
D^0_{CP}\phi.$}

To extract the CKM-phase $\gamma$, it was necessary to assume knowledge on
a ratio of unmixed amplitudes, such as
$\Gamma (B_s \rightarrow D^-_s K^+ ) /\Gamma (B_s\rightarrow D^-_s \pi^+.)$
Time-dependent studies of untagged data samples of $\overline D^0 \phi ,
D^0 \phi ,D^0_{CP}\phi$ extract $\gamma\; (=-\phi )$ without any
assumptions, except the validity of the CKM model.
They even determine $|\lambda |$ and the strong phase-difference $\Delta$.
Denote by $\eta \;(+1 \;{\text or}\;-1)$ the CP-parity of
$D^0_{CP}$. Thus the CP-parity of the whole $B_s$-mode $D^0_{CP} \phi$
is $(-\eta )$.
 The time dependences
determine, respectively,
\begin{equation}
\frac{Re\lambda}{1+|\lambda |^2} \;, \;\;\frac{Re\overline{\lambda}}{1+
|\lambda |^2} \;, \;\;\frac{Re\lambda_{\eta}}{1+|\lambda_{\eta} |^2} \;,
\end{equation}
where
$$
\lambda \;\equiv \;\frac{q}{p}\;
\frac{\langle \overline D^0 \phi |\overline B_s \rangle }{\langle \overline D^0
\phi |B_s \rangle } = |\lambda | \;e^{i(\phi +\Delta )} \;,
$$
$$
\overline{\lambda} \;\equiv \;\frac{p}{q}\;
\frac{\langle D^0 \phi | B_s \rangle }{\langle D^0 \phi |\overline B_s \rangle
} = |\lambda |
\;e^{i(-\phi +\Delta)} \;,
$$
$$
\overline{\lambda} = \lambda\; e^{-2i\phi} \;,
$$
\begin{equation}
\lambda_{\eta} \;\equiv \;\frac{q}{p}\;
\frac{\langle D^0_{CP} \phi |\overline B_s \rangle }{\langle D^0_{CP}\phi |B_s
\rangle } =
\frac{\eta\; \lambda \;-\; 1 }{\eta -\overline \lambda}\;.
\end{equation}
The three unknowns $|\lambda |, \phi$ and $\Delta$ can be
determined from the three measurables, Eq.~(5.33).
The magnitude of the interference term $|\lambda |$ could be obtained
alternatively by using theory on the ratio [see Eq.~(5.26)],
\begin{equation}
\frac{\Gamma (B_s \rightarrow \overline{D}^0 \phi )}{\Gamma (B_s \rightarrow
\overline{D}^0 \overline{K}^{*0})} \;.
\end{equation}
We suspect, however, that theory cannot predict as reliably this ratio of
rates, because rescattering effects may be more pronounced for the
color-suppressed modes than for the color-allowed ones.  A comparison of the
two determinations of $|\lambda |$ therefore probes rescattering effects.

\subsubsection{CP Violation}

Time-dependences of untagged $B_s$ modes governed by $\bar b\rightarrow
\bar c u\bar s$ could show sizable CP violating effects.  CP invariance demands
that
\begin{equation}
\Gamma \left[f\left(t\right)\right] =\Gamma\left[ \overline f
\left(t\right)\right] \;,
\end{equation}
or equivalently,

\begin{equation}
Re\lambda =Re\overline \lambda \Longleftrightarrow \cos (\phi +\Delta )=\cos
(-\phi +\Delta )\;.
\end{equation}
Thus CP-violation will be more pronounced for modes where
$\Delta$ is more sizable. We expect the color-suppressed modes
to show larger CP-violating effects than the color-allowed modes, where
$\Delta$ is expected to be smaller.

It is very important to realize that the $B_s$ meson harbors possibly
large CP-violating effects, for which one is not required to distinguish an
initial
$B_s$ and $\overline{B}_s$. Such CP-violating effects are the
time-dependent or time-integrated asymmetries,
\begin{equation}
a(t) \equiv \frac{\Gamma [f(t)] - \Gamma [ \overline f   (t)]}
{\Gamma [f(t)] + \Gamma [ \overline f   (t)]}\; ,
\end{equation}
\begin{equation}
A(t_0 )\equiv \frac{\int^\infty_{t_0} \;dt \;\bigg\{\Gamma \left[f\left(t
\right)\right] -\Gamma \left[ \overline f
\left(t\right)\right]\bigg\}}
{\int^\infty_{t_0} \;dt\;\bigg\{\Gamma\left[f\left(t\right)\right] +
\Gamma \left[  \overline f  \left(t\right)\right]\bigg\}}\; .
\end{equation}
Eqs.~(5.21) and (5.38) yield
\begin{equation}
a(t) = \frac{-2|\lambda |\sin\phi\;\sin\Delta\;\tanh\left(\frac{\Delta
\Gamma t}{2}\right)}
{1+|\lambda |^2 +2|\lambda |\cos\phi\;\cos\Delta\;  \tanh\left(\frac{\Delta
\Gamma t}{2}\right)} \;.
\end{equation}
In the limit
\begin{equation}
\lim_{t\rightarrow\infty} \;\tanh \left(\frac{\Delta\Gamma \; t}{2}\right) =-1
\;,
\end{equation}
which is satisfied in practice for
\begin{equation}
t \;\raisebox{-.4ex}{\rlap{$\sim$}} \raisebox{.4ex}{$>$}\; \frac{2}{\Delta
\Gamma} \;,
\end{equation}
one finds
\begin{equation}
\lim_{t\rightarrow\infty} \;a(t) =\frac{2|\lambda |\sin\phi\;\sin\Delta}
{1+|\lambda |^2 -2|\lambda |\cos\phi\;\cos\Delta} \;.
\end{equation}
To demonstrate that large CP violating effects are possible, proper decay times
greater than about $2/\Delta\Gamma$ are used.  Clearly, to optimize observation
of CP violation and the extraction of CKM-phases we recommend to
\underline{always} use all accessible proper times.
Representative values for modes governed by $\overline b\rightarrow \overline c
u \overline s$---such as $\overline{D^0}\phi,\;\; D^{(*)-}_s K^{(*)+}$---would
be
\begin{equation}
|\lambda |=\frac{1}{3} \;,\;\;\sin\phi = -0.8 \;, \;\;\cos\phi =0.6\;.
\end{equation}
For a large phase difference $\Delta =30^0$, more relevant for
$\overline{D^0}\phi$, we find
\begin{equation}
a(\infty ) =-0.35 \;,
\end{equation}
whereas for $\Delta =5^0$, probably more in line for
$D^{(*)-}_s K^{(*)+}$, we find
\begin{equation}
a(\infty )=  - 0.065 \;.
\end{equation}
Even larger asymmetries can be envisioned.
Such asymmetries would not be diluted by the
many tagging inefficiencies and dilution effects encountered in
asymmetries that require separation of $B^0$ and $\overline{B}^0 $ mesons.
Time is the tag here. By waiting long enough, the faster decaying of the two
$B_s$ mass-eigenstates has vanished. What is seen is the remnant of the slower
decaying $B_s$ mass-eigenstate.

We lose lots of statistics because we study decays at about
$2/\Delta\Gamma\approx 7$ lifetimes or more. But such long lived $B$'s may
harbor
sizable effects, without any additional dilutions. One cannot
but be struck by the comparison to the $K_L$ and $K_S$ mesons. Whereas
there is no loss in statistics in separating $K_L$ out from $K^0$, because
$\tau_{K_L} \approx 600 \;\tau_{K_S} $, the involved CP-violating effects are
minuscule and very hard to interpret in terms of the fundamental
CKM-parameters.
In contrast, separating $B_H$ out from $B_s$ requires large statistics,
because times $t\;\raisebox{-.4ex}{\rlap{$\sim$}} \raisebox{.4ex}{$>$}\;
\frac{2}{\Delta\Gamma}$ are used, but the CP-violating effects can
be significant and the relevant CKM-parameters can be extracted.

\subsubsection{Resonance Effects}

Studies of $B$
modes where several resonances contribute to the final state may enhance CP
violating effects as discussed by Atwood et al.. They applied
their method to final states governed by the $b\rightarrow s\gamma , d\gamma$
\cite{atwoodsoni} transitions and by the $b \rightarrow s\overline D^0,\; s
D^0,\; s D^0_{CP}$
\cite{atwood} transitions.  Sizable CP violating observables can be constructed
for $B_s$ modes such as
$D^{(*)-}_s (K\pi ), D^{(*)-}_s (K^* \pi ), D^{(*)-}_s
(K\rho ), D^{(*)-}_s (K \pi \pi)$ where the particles in parentheses originate
from several
interfering kaon resonances. Those modes also extract the CKM-phase
$\gamma$ and may eliminate a two-fold ambiguity in the determination
of $\sin\gamma$. A detailed study is underway~\cite{adsy}.

To summarize, this subsection described the extraction
of the CKM-phase $\gamma$
from time-dependences of untagged $B_s$ modes governed by
$\bar b\rightarrow \bar c u\bar s$. CP violating effects may be sizable and
are enhanced by resonance effects.

\subsection{Modes with Several CKM-Contributions}

Consider first flavor-specific modes $g$ where several CKM-combinations
contribute to the unmixed decay-amplitude,
\begin{equation}
B_s \rightarrow g, \;\;\overline B_s \not\rightarrow g , \;\;\lambda =
\overline
\lambda =0 \;,
\end{equation}
for example $K^{(*)-} \pi^+ , \;K^- (\pi^+ \pi^+\pi^- ), \;J/\psi \overline
K^{*0}
(\rightarrow K^- \pi^+ ), \;J/\psi K^- \pi^+, D_s^{(*)-}D^{(*)+}$.
The untagged time-evolution is given by,
\begin{equation}
\Gamma \left[ g\left(t\right)\right] = \frac{\Gamma (B^0 \rightarrow g )}{2}
\;\bigg\{ e^{-\Gamma_L t} + e^{-\Gamma_H t}\bigg\}\;,
\end{equation}
\begin{equation}
\Gamma \left[ \overline g\left(t\right)\right] = \frac{\Gamma (\overline B^0
\rightarrow \overline g )}{2}
\;\bigg\{ e^{-\Gamma_L t} + e^{-\Gamma_H t}\bigg\}\;.
\end{equation}
The modes $g$ may show direct CP violation \cite{bss,bjarlskog}, where the
CP-violating asymmetry is
\begin{equation}
A_g \equiv \frac{\Gamma (B^0 \rightarrow g) - \Gamma (\overline B^0 \rightarrow
\overline g )}
{\Gamma (B^0 \rightarrow g) + \Gamma (\overline B^0 \rightarrow
\overline g )} \;.
\end{equation}
The same asymmetry can be seen as either a time-dependent or a time-integrated
effect,
\begin{equation}
A_g  =  \frac{\Gamma \left[ g\left(t\right)\right] - \Gamma \left[ \overline g
\left(t\right)\right]}
{\Gamma \left[ g\left(t\right)\right] + \Gamma \left[ \overline g\left(t\right)
\right]} =
\frac{\int^\infty_{t_0} \;dt \;\bigg\{ \Gamma \left[ g\left(t\right)\right]
-\Gamma\left[ \overline g \left(t\right)\right]\bigg\}}
{\int^\infty_{t_0} \;dt \;\bigg\{\Gamma\left[g\left(t\right)\right]
+\Gamma [\overline g\left(t\right)]\bigg\}} \;.
\end{equation}

Modes common to $B_s$ and $\overline B_s$ where several CKM-combinations
contribute to $B_s \rightarrow f$ may show direct CP violation
$[\Gamma (B_s \rightarrow f)\neq \Gamma (\overline B_s \rightarrow
\overline f)]$ as well as CP violation due to mixing.
CP invariance demands that
$\Gamma [f(t)] =\Gamma [\overline f(t)]$. The
time-evolution of untagged modes $f$ and $\overline f$ allows one to
disentangle partially the various CP-violating effects.
The $B_s$ modes $K^+ K^- , \phi K_S , \rho^0
K_S , D^0_{CP} \phi , J/\psi K_S, \phi\phi, \rho^0 \phi$, etc. all serve as
examples.

\subsection{Measuring the Width Difference}

After having derived the time-dependent formulae in previous subsections,
we are now in a position to list several suggestions for determing
$\Delta\Gamma$. A detailed feasibility study will be presented elsewhere
\cite{feasibility}. All the methods may be combined to optimize
the determination.

The first two methods use the important observation that the
average $B_s$ width $\Gamma$ is in fact already known \cite{bigi,bigietal}.
Table I shows the predicted \cite{bigi,bigietal} and measured
\cite{forty} ratios of lifetimes of $b$-flavored hadrons.

Refs.~\cite{bigi,bigietal} claim the following.
The $B^-$ lifetime is predicted to be longer than the $B_d$ lifetime due to
Pauli interference.  For the neutral $B$ mesons, the $W$-annihilation
amplitudes ($b \overline d \to c \overline u,
b \overline s \to c \overline c$)
are helicity suppressed and unimportant numerically,
which yields same lifetimes for the average $B_s$ and $B_d$ mesons. The
$\Lambda_b$ lifetime prediction still requires a careful theoretical analysis
but
it is claimed that
\begin{equation}
0.9 \;\raisebox{-.4ex}{\rlap{$\sim$}} \raisebox{.4ex}{$<$}\;
\frac{\tau (\Lambda_b)}{\tau (B_d)} \;<\;1 \;.
\end{equation}

Refs. \cite{bigi,bigietal} must be critically re-evaluated, however, because
they obtain a too large semileptonic branching ratio and
too small an inclusive width for the $b\rightarrow c\bar cs$ transition
in $B$ decays \cite{baffle,fwd,voloshin,bagan}.  Further, the $W$-annihilation
amplitude interferes with different spectator decays.  It interferes with the
spectator decay
$b \to c \bar u d, b \to c \bar c s$ for the $\overline B_d,\; \overline B_s$,
respectively. We believe
that the $b \to c \bar c s$ transition is the least understood theoretically.
A detailed study, which estimates how different the $B_d$, $B_s$ and other
$b$-hadron lifetimes can be, would be useful.  Because such a critical
re-evaluation is still lacking, this subsection uses the predictions
of Bigi et al. \cite{bigi,bigietal}, with the understanding that their
estimates require refinement.

The average decay-width $\Gamma$ of $B_s$ could be determined essentially from
a one parameter  fit $\exp (-\Gamma t)$ of the time-evolution of the untagged,
flavor-specific data sample~\cite{feasibility}.  It could be deduced
alternatively from the measured lifetimes of other $b$-species.  For instance,
the $B_d$ and average $B_s$  [$\overline \tau(B_s) \equiv 1/\Gamma$] lifetimes
are claimed to be equal to
excellent accuracy~\cite{Bdlifetime}. Thus the average decay-width $\Gamma$ of
$B_s$ is measured.
The width $\Gamma$ can also be obtained from inclusive $b$
lifetime measurements.
Denote by $T$ a particle, collection of particles, or event topology,  which
characterizes $b$-decay.  Examples for $T$ are detached $J/\psi$, primary
leptons
(i.e., leptons in $b\rightarrow c\ell$ processes) with an impact
parameter, such primary leptons in coincidence with detached vertices, or
detached multi-prong vertices,
where the whole event is consistent with being a $b$-decay.

A single exponential fit of the proper (multi-exponential) time distribution of
this inclusive $b$-data sample
determines the ``average" $b$-lifetime $\tau (b)$,

\begin{eqnarray}
e^{-t/\tau (b)} & \sim & p_d \;R(B_d \rightarrow TX )\; e^{-t/\tau (B_d)} +
\nonumber \\
&   & p_u \;R(B_u \rightarrow TX ) \;e^{-t/\tau (B_u)} + \nonumber \\
&   & p_s \;R(B_s \rightarrow TX) \;S(t) + \nonumber \\
&   & p_{\Lambda_b} \;R(\Lambda_b\rightarrow TX) \;e^{-t/\tau (\Lambda_b )} \;.
\end{eqnarray}
The production fractions for $\overline B_d,\; B_u^-,\;
\overline B_s,\;\Lambda_b$ are assumed to be~\cite{bbhep}
\begin{equation}
p_d \;: \;p_u \;: \;p_s \;: \;p_{\Lambda_b} \approx
0.375 \;:\; 0.375 \;: \;0.15 \;: \;0.10 \;.
\end{equation}
The inclusive yield of $T$ in $b$-hadron decay is defined as
\begin{equation}
R(H_b \rightarrow TX) \equiv B(H_b\rightarrow TX) + B
(\overline{H}_b \rightarrow TX) \;,
\end{equation}
for $H_b = \overline B_d , B_u^- , \overline B_s , \Lambda_b$.
The function $S(t)$ depends on which inclusive data sample is used.  For
flavor-specific $T$ [such as $\ell^\pm X$]
\begin{equation}
S(t) = \frac{e^{-\Gamma_L t} +e^{-\Gamma_H t}}{2}\;,
\end{equation}
whereas for flavor-nonspecific $T$ [such as $J/\psi X$]
\begin{equation}
S(t) =  e^{-\Gamma_L t} \;.
\end{equation}
Eq.~(5.57) assumes that the inclusive flavor-nonspecific $T$ production in
$B_s$ decays [such as the prominent $J/\psi$] is dominated by CP-even modes.

It is instructive to approximate $\tau (b)$ for an inclusive flavor-specific
data sample $T$ as
\begin{equation}
\tau (b) \approx [p_d \;/\; \tau (B_d) + p_u \;/\;\tau (B_u) + p_s
\;/\;\overline \tau (B_s)
+ p_{\Lambda_b} \;/\;\tau (\Lambda_b)]^{-1} \;.
\end{equation}
This approximation uses the observation and prediction of small differences in
separate $b$-hadron
lifetimes and further assumes equal inclusive yields of $T$ in all
$H_b$ decays.
Using Table I and the assumed
specific $b$-hadron production
fractions, we get from Eq.~(5.58)
\begin{equation}
\tau (b) =\tau (B_d ) \;\left[1\pm {\cal O} \left(0.01\right) \right]\;.
\end{equation}
The truly inclusive $b$-lifetime measures essentially the $B_d$ lifetime,
which in turn is essentially the average $B_s$ lifetime. The
average width $\Gamma$ of $B_s$ is thus known
\begin{equation}
\Gamma \approx 1/\tau (b) \;.
\end{equation}
In summary, $\Gamma$ is essentially known from either a single parameter fit of
the untagged, flavor-specific $B_s$ data sample, or from lifetime measurements
of either $B_d$'s or inclusive $b$-decays.
We are now ready to discuss several methods for extracting $\Delta\Gamma$.

\begin{flushleft}
\underline{Method 1}
\end{flushleft}

The proper time-dependence of untagged flavor-specific modes
of $B_s$ is given by
\begin{equation}
e^{-\left(\Gamma + \frac{\Delta\Gamma}{2}\right)t} + e^{-\left(\Gamma -
\frac{\Delta\Gamma}{2}\right)t} \;.
\end{equation}
The average width $\Gamma$ is known and a one parameter
fit of the measured time-dependence determines $\Delta\Gamma$.
More than 200 flavor specific $B_s$-events have already been recorded at
LEP and CDF.  This method may be rather effective.

\begin{flushleft}
\underline{Method 2}
\end{flushleft}

The CP-even [CP-odd] $B_s$ modes driven by $b\rightarrow c\bar cs$ are
governed by a single exponential decay law
\begin{equation}
e^{-\Gamma_L t} = e^{-\left(\Gamma -\frac{\Delta\Gamma}{2}\right)t}
\left[e^{-\Gamma_H t} = e^{-\left(\Gamma
+\frac{\Delta\Gamma}{2}\right)t}\right] \;.
\end{equation}
Combining this determination of $\Gamma_L\;\;  [\Gamma_H ]$
with the known $\Gamma$ measures $\Delta\Gamma$.

For Methods 1 and 2, we may wish to parametrize our ignorance as to the exact
value of $\Gamma$ by a small parameter $\epsilon$,
\begin{equation}
\Gamma \to \Gamma\; +\; \epsilon.
\end{equation}
A two parameter fit would extract both $\Delta\Gamma$ and $\epsilon$.  In
contrast to Methods 1--2, Methods 3--7 do not assume knowledge of $\Gamma$.

\begin{flushleft}
\underline{Method 3}
\end{flushleft}

This is basically the method advocated by Bigi et al. \cite{bigi,bigietal},
which we reviewed and refined in previous subsections. The time-evolutions of
untagged, flavor-specific modes and of CP-even modes of $B_s$ governed by
$b\rightarrow c\bar cs$
are given by Eqs.~(5.4) and (5.62), respectively.
The CP-even modes determine $\Gamma_L$.
A one-parameter fit of the time-evolution of the untagged, flavor-specific
modes
determines $\Gamma_H$, because $\Gamma_L$ has been measured.
Of course, the exponential decay law of the CP-odd $B_s$ modes driven by
$b\rightarrow
c\bar cs$ can be used as a consistency check and must be governed by
$\Gamma_H$.

\begin{flushleft}
\underline{Method 4}
\end{flushleft}

The time-evolution of the CP-even and CP-odd eigenmodes driven by the
$b\rightarrow c\bar c s$ transition are governed by $\Gamma_+ = \Gamma_L$ and
$\Gamma_- = \Gamma_H$, respectively.  A time-dependent study of untagged
CP-even and CP-odd modes measures the width difference.  The CP-even modes are
expected to dominate over the CP-odd ones, and are probably also easier to
detect.  To increase usable data sets with definite CP, Ref.~\cite{bigirosner}
suggested employing angular correlations~\cite{kkps,dqstl} to decompose modes
that are mixtures of CP-even and CP-odd parities [such as $J/\psi
\phi,D_s^{*+}D_s^{*-},J/\psi \phi\rho^0$] into definite CP-components.

\begin{flushleft}
\underline{Method 5}
\end{flushleft}

Any mode governed by $b\rightarrow c\bar c s$, which is a mixture of CP-even
and CP-odd parities [for example, $J/\psi \phi,D_s^{*+}D_s^{*-},J/\psi
\phi\rho^0$], allows the extraction of both $\Gamma_H$ and $\Gamma_L$.  This
has been discussed in Ref.~\cite{bigirosner} by decomposing such modes into
CP-even and CP-odd components and studying their different decay laws.  The
extraction of $\Delta\Gamma$ from such modes is optimized however by a complete
study of angular correlations~\cite{dqstl} combined with other relevant
techniques (such as Dalitz plots, etc.), which we advocate.  Time-evolutions of
interference terms will add valuable information on top of the time-dependences
of the definite CP-components.  Such a study truly optimizes the determination
of $\Delta\Gamma$ from modes which are admixtures of CP-even and CP-odd
parities.

\begin{flushleft}
\underline{Method 6}
\end{flushleft}

For a small width difference, one may be able to determine $\Delta\Gamma$ from
CP violating effects with untagged $B_s$ data samples, such as the asymmetries
discussed in Eqs.~(5.38)-(5.40).  A time-dependent fit may be able to determine
the argument of $\tanh$ and thus $\Delta\Gamma$, see for instance Eq.~(5.40).
The determination is facilitated by knowing $|\lambda|, \phi$ and $\Delta$.
$|\lambda|$ can be obtained as discussed in Section (V.C). The weak phase
$\phi$ will be well known from other techniques by the time such a measurement
of $\Delta\Gamma$ becomes feasible.  As for the final-state phase $\Delta$, it
is calculable for example for $B_s$ modes where several resonances contribute
to the final state, such as
$D^{(*)-}_s (K\pi ), D^{(*)-}_s (K^* \pi ), D^{(*)-}_s
(K\rho ), D^{(*)-}_s (K \pi \pi)$.

\begin{flushleft}
\underline{Method 7}
\end{flushleft}

There exist $B_s$ modes with time-evolutions that depend on both the
sum and the differences of the two exponents,
\begin{equation}
e^{-\Gamma_L t} \;\pm \;e^{-\Gamma_H t} \;.
\end{equation}
A fit to these time-evolutions determines both $\Gamma_L$ and $\Gamma_H$
\cite{feasibility}. Within the CKM model, such modes are CKM-suppressed
and probably not competitive with other methods.
However, if the CKM model is broken and CP-eigenmodes of $B_s$ driven by
$b\rightarrow c\bar cs$ show two different exponential decay laws, then
this method is one possible way to measure both widths.

Those are then some possible ways for extracting $\Delta\Gamma$. We wish
to conclude this section with a suggestion of how to enrich a $B$ data
sample with $B_s$ mesons.  The key is a $\phi$-trigger \cite{phitrigger}.
The $\phi$ is seen in the $K^+K^-$ mode about 50\% of the time.
This mode occurs close to threshold. For energetic $\phi$'s, the
two charged kaons have roughly equal momenta and go in similar directions.
This may simplify triggering on $\phi$'s.
The inclusive $\phi$ yield is about 20\% in $D_s$ decays,
whereas it is roughly an order magnitude less in other charmed hadron decays
\cite{pdg,distinguish}.
Thus, $\phi$'s discriminate well between $D_s$ and other charmed hadrons.
Further, it is believed that the inclusive yield of $D_s$ in $B_s$ decays is
quite enhanced over that in $B$ decays. Inclusive $b$-decays with a
$D_s$ in the final state enrich the $B_s$ content of that $b$-sample.
In fact, the DELPHI collaboration used $\phi\ell X$ modes as an enriched
$B_s$-sample
and extracted an average $B_s$ lifetime from it \cite{delphi}.
We hope to see flavor-specific modes like $B_s\rightarrow\phi\ell X$ being
used both at $e^+e^-$ and $p\bar p$ colliders to extract not only
the average $B_s$ lifetime, but the width difference $\Delta\Gamma$ as well.

\section{Conclusions}

Theoretical predictions for a sizable lifetime difference between the light
and heavy $B_s$ mass-eigenstates have existed for many years
\cite{box,shifman,dpt,dthesis,dunietzwidth,aleksan,bigi,bigietal}.
The observation of a non-vanishing $\Delta\Gamma$ would prove the existence of
$B_s - \overline{B}_s$
mixing. How could such a width difference be determined experimentally?
To that effect, we considered the time-evolution of untagged data samples
of $B_s$ mesons. We found that the rapid oscillatory behavior
$\Delta mt$ cancels in all untagged samples, provided that
$\left|\frac{q}{p}\right| \approx 1$ which is satisfied to
${\cal O}(10^{-3}-10^{-4})$ within the CKM-model.
The time-evolution of untagged data samples are governed solely by
the two exponential falloffs, $e^{-\Gamma_L t}$ and $e^{-\Gamma_H t}$, which
enables $\Delta\Gamma$ to be measured in several ways; see Section (V.E).
The exponentials are much more slowly varying functions of proper time than the
rapid $\Delta mt$-oscillations.
This allows us to conduct feasibility studies with presently existing
technology~\cite{feasibility}.

Once the two widths are known and found to differ, CP violation can be
seen with untagged, time-evolved data samples of $B_s$.
CP invariance demands that modes of $B_s$ with definite CP-parity (i.e., that
are CP eigenstates) time-evolve with a single exponential. Thus if the
time-evolution of CP-eigenstates, such as $\rho^0 K_S , D^0_{CP}\phi ,K^+K^- ,$
 has two non-vanishing exponential falloffs, CP violation has
been demonstrated. The demonstration can clearly already occur for
\underline{untagged} data samples.

The time-evolution of the \underline{untagged} $\rho^0 K_S$ data sample
is not only useful in observing CP violation but \underline{even extracts}
$\cos 2\gamma$, when penguin contributions are neglected. Those penguin
effects could be sizeable however, and thus we discussed next the
extraction of the unitarity angle $\gamma$ from time-dependences of
untagged $B_s$ data samples governed by $\bar b \rightarrow \bar c u
\bar s$. Penguin amplitudes are absent. The time-evolution of untagged, for
instance,
$D_s^{(*)\pm} K^{(*)\mp}, \stackrel{(-)}{D^0} \phi \;B_s$-modes measure
$\cos (\gamma +\Delta )$ and $\cos (-\gamma +\Delta )$, with the overall
normalization being determined from flavor specific modes, such as
$D^{(*)\pm}_s
\pi^\mp , D^{(*)\pm}_s \ell\nu$, $D^0 K^{*0} ,\overline{D}^0\overline{K}^{*0}$.
A two-fold ambiguity in $\sin\gamma$ can be resolved, and both $\sin\gamma$
and $sin^2 \Delta$ are extracted.

The above extraction of the phases $\gamma$ and $\Delta$ involves the
factorization assumption to determine the
normalization. For those who object to this
assumption, there exist a series of measurements that extracts $\gamma$
without any theoretical input. The time-evolutions of the untagged data
samples $D^0 \phi , \overline{D}^0 \phi$ and $D^0_{CP}\phi$
determine $|\lambda |,\gamma$ and $\Delta$ without any theory.
The measured $|\lambda |$ can then be cross-checked with its measurement
involving
some theory input, which allows insights into rescattering effects
of color-suppressed processes.  Sizable CP-violating effects can occur
with those untagged data samples for large enough proper times.
A detailed study is underway which addresses the feasibility of all the
above-mentioned measurements for a generic detector \cite{feasibility}.

If such measurements turn out to be feasible, then arguments can be made in
favor of $p \overline p$ and $e^+ e^-$ experiments versus $pp$ or $ep$
experiments.  The former experiments have a charge-symmetric initial state
which allows trivial recording of untagged $B_s$ data samples, in contrast to
the latter experiments.

The ramifications of a large width difference for the $B_s$-meson
are far reaching. The ratio $\frac{\Delta m}{\Delta\Gamma}$ involves no
CKM-combination, only a QCD uncertainty. If a careful study finds that
this ratio can be rather well estimated, then, by observing
either $\Delta\Gamma$ or $\Delta m$ first, the other difference will be known
too
(within the CKM-model).
The ratio $\frac{\Delta m}{\Delta\Gamma}$ may turn out to be an important
Standard
Model constraint.  Second, a large width difference will prove that
$B(b\rightarrow c\bar cs$) is sizable and would solve the
so-called puzzle of the number of charmed hadrons per $B$-meson. It would show
that experimentalists simply have not taken the detection efficiencies of
exotic
charmed hadron yields in $B$ decay and absolute branching fractions of charmed
hadron decays carefully into account.
In so doing, they created a spurious puzzle \cite{baffle,fwd,voloshin,bagan}.
Third, one will not be
allowed to speak about branching fractions of $B_s^0 \rightarrow f$,
but only about $B(B_{H,L}\rightarrow f)$.

An analogy to the neutral kaons is instructive. The $K_L$ lives about
600 times as long as a $K_S$, thus a $K^0$ or $\overline{K}^0$ is
essentially a $K_L$ after a few $K_S$ lifetimes, without having lost almost
any $K_L$. The CP violating effects are tiny and the extraction of the
fundamental CKM-parameters is messy  because of large uncertainties in
strong matrix elements.

 In contrast, the $B_s$ meson has comparable
widths for the heavy and light mass-eigenstates. They differ at the
(20-30)\% level.
To guarantee a pure data sample of $B_H$ requires one to go out to about seven
$B_s$
lifetimes, costing tremendously in statistics.
But then many exciting measurements become feasible, because the $b$
proceeds to decay through several quark transitions into many possible
final states.
Sizable CP violating effects and the clean extraction of
fundamental CKM-parameters may be possible with untagged data samples of
$B_s$ mesons. (Clearly, to optimize the measurements not only pure $B_H$ data
samples but rather all available proper times better be used.)  As in the case
of neutral kaons, time plays the role
of the ``tag". Many more measurements can be contemplated than what is reported
here, once $\Delta\Gamma$ is found to be nonvanishing.

\section{Acknowledgements}

This note is a continuation of my studies with R.G. Sachs, J.L. Rosner,
J.D.~Bjorken, B. Winstein, J. Cronin and all my other colleagues while I was a
graduate student at the University of Chicago.  We are grateful to them.  We
also thank Joe Incandela, Eric Kajfasz, Rick Snider and Dave Stuart for
informative discussions concerning feasibilities, and Gerhard Buchalla for
discussions concerning QCD corrections. We thank Lois Deringer for help in
typing this
manuscript.  This work was supported by the Department of Energy, Contract
No.~DE-AC02-76CHO3000.

\begin{table}
\caption{Predicted [16,17]
%bigi,bigietal
and measured [9]
%forty
lifetime ratios of $b$-flavored hadrons.}
\begin{tabular}{|l|c|c|}
& Prediction & Data \\
\hline
$\tau (B^- )/\tau (B_d)$ & $1+0.05 \left(\frac{f_B}{200
\;\text{MeV}}\right)^2
\left[1 \;\pm \; {\cal O} \left(10\%\right)\right]$ & $1.01 \;\pm
\;0.09$ \\
\hline
$\overline{\tau} (B_s )/\tau (B_d )$ & 1 $\;\pm\; {\cal O}$ (0.01) &
$0.98 \;\pm
\;0.12$ \\
\hline
$\tau (\Lambda_b ) /\tau (B_d )$ & $\sim$ 0.9 & 0.71 $\;\pm\;$ 0.10
\\
\end{tabular}
\end{table}


\begin{references}

\bibitem{bscdf}
CDF Collaboration, Fermilab report, FERMILAB-CONF-94/138-E.

\bibitem{bslep}
J. Cuevas (DELPHI Collaboration);
D. Karlen (OPAL Collaboration);
V. Sharma (ALEPH Collaboration), talks given at DPF'94, University of New
Mexico,
Albuquerque, New Mexico, August, 1994.

\bibitem{opal}
P.D.~Acton et al. (OPAL Collaboration), Phys. Lett. {\bf B312}, 501 (1993).

\bibitem{aleph}
D. Buskulic et al. (ALEPH Collaboration), Phys. Lett. {\bf B322}, 275 (1994).

\bibitem{delphi}
P.~Abreu et al. (DELPHI Collaboration), Z. Phys. {\bf C61}, 407 (1994).

\bibitem{ckm}
N. Cabibbo, Phys. Rev. Lett. {\bf 10}, 531 (1963);
M. Kobayashi and T. Maskawa, Prog. Theor. Phys. {\bf 49}, 652 (1973).

\bibitem{paschoszacher}
L.~Wolfenstein, Nucl. Phys. {\bf B246}, 45 (1984);
N.W. Reay, in Proceedings of the SSC Fixed Target Workshop, The Woodlands,
Texas, 1984, eds. P.~McIntyre et al., p.~53;
E.A. Paschos and R.A. Zacher, Z. Phys. {\bf C28}, 521 (1985).

\bibitem{dr}
I.~Dunietz and J.L.~Rosner, Phys. Rev. {\bf D34}, 1404 (1986).

\bibitem{forty}
R.~Forty, CERN report, CERN-PPE/94-144, September 1994, invited talk at the XIV
International Conference on Physics in Collision, Tallahassee, June 15--17,
1994.

\bibitem{box}
J.S. Hagelin, Nucl. Phys. {\bf B193}, 123 (1981);
E. Franco, M. Lusignoli, A. Pugliese, Nucl. Phys. {\bf B194}, 403 (1982);
A.J. Buras, W. Slominski, and H. Steger, Nucl. Phys. {\bf B245}, 369 (1984).

\bibitem{shifman}
M. B. Voloshin, N. G. Uraltsev, V. A. Khoze and M.A. Shifman, Sov. J.
Nucl. Phys. {\bf 46}, 112 (1987).

\bibitem{dpt}
A. Datta, E.A. Paschos and U. T\"urke, Phys. Lett. {\bf B196}, 382
(1987).

\bibitem{dthesis}
I.~Dunietz, Ph.D. Thesis, Ann. Phys. {\bf 184}, 350 (1988), and
references therein.

\bibitem{dunietzwidth}
I.~Dunietz, contribution to Workshop on High Sensitivity Beauty
Physics, Fermilab, Batavia,
IL, Nov 11-14, 1987, editors A.J. Slaughter, N. Lockyer, and M.
Schmidt, p. 229.

\bibitem{aleksan}
R. Aleksan, A. Le Yaouanc, L. Oliver, O. P\`ene and J.-C.
Raynal, Phys. Lett. {\bf B316}, 567 (1993).

\bibitem{bigi}
I.I. Bigi, CERN Report, CERN-TH.7282/94, May 1994;
CERN Report, CERN-TH.7050/93, October 1993.

\bibitem{bigietal}
I. Bigi, B. Blok, M. Shifman, N. Uraltsev, A. Vainshtein, CERN Report,
CERN-TH.7132/94, January 1994, in the second edition of the book {\it B
decays}, p. 132, ed. S. Stone, World Scientific, 1994, and references therein.

\bibitem{cpreviewBigi}
I. I.  Bigi, V.A. Khoze, N.G. Uraltsev, and A.I. Sanda, in {\it CP Violation},
edited by C. Jarlskog (World Scientific, Singapore, 1989), p. 175.

\bibitem{cpreviewNir}
Y. Nir, SLAC Report, SLAC-PUB-5874, September 1992,
Lectures given at 20th Annual SLAC Summer Institute on Particle Physics: The
Third Family and the Physics of Flavor.

\bibitem{cpreviewMcDonald}
K.T. McDonald, Princeton Report, Princeton/HEP/92-09, September 1992
(unpublished).

\bibitem{xs}
D.J.~Ritchie et al., p. 357;
J.E.~Skarha and A.B.~Wicklund, p. 361;
T.H.~Burnett, p. 367;
T.J.~Lawry et al., p. 371;
X.~Lou, p. 373;
C.~Baltay et al., p. 377;
K.~Johns, p. 383;
all in Proceedings of the Workshop on $B$ Physics at Hadron Accelerators,
Snowmass, Colorado, June 21 - July 2, 1993, ed. by Patricia~McBride and
C.~Shekhar~Mishra.

\bibitem{xsnir}
A. Ali and D. London, J. Phys. {\bf G19}, 1069 (1993);
CERN Report, CERN-TH-7398-94, Aug 1994;
Y. Nir, Phys. Lett. {\bf B327}, 85 (1994).

\bibitem{sharma}
V. Sharma (ALEPH Collaboration), plenary talk given at DPF'94, University of
New Mexico,
Albuquerque, New Mexico, August, 1994.

\bibitem{baffle}
I.I.~Bigi, B.~Blok, M.A.~Shifman and A.~Vainshtein, Phys.\ Lett.\
{\bf B323}, 408 (1994).

\bibitem{fwd}
A.F.~Falk, M.B.~Wise, and I.~Dunietz, Fermilab Report
No.~FERMILAB--PUB--94--106--T, to appear in Phys.\ Rev. {\bf D}.

\bibitem{voloshin}
M.B. Voloshin, Minnesota Report, TPI-MINN-94-35-T, September 1994 (hep-ph -
9409391).

\bibitem{bagan}
E. Bagan, P. Ball, V.M. Braun, and P. Gosdzinsky, DESY Report, DESY 94-172,
September 1994.

\bibitem{feasibility}
I. Dunietz et al., Fermilab Report No.~FERMILAB--PUB--94/362--T, in
preparation.
%F. DeJongh (???), I.~Dunietz,
%J.~Incandela, E.~Kajfasz, J.~Kroll (??),
%M.~Paulini (?), R.~Snider, P. Sphicas and D.~Stuart,

\bibitem{confusion}
One must distinguish between $\Gamma_\pm$ and $\Gamma_\pm (b\rightarrow c\bar c
s)$.  In the absence of CP violation $\Gamma_\pm$ denotes the \underline{total}
decay widths of the light and heavy $B_s$ eigenstates.  In contrast $\Gamma_\pm
(b\rightarrow c\bar c s)$ denotes the CP-even and CP-odd $B_s$-rates governed
by the $b\rightarrow c\bar c s$ transition alone.

\bibitem{inamiLim}
T. Inami and C.S. Lim, Prog. Theor. Phys. {\bf 65}, 297 (1981); ERRATUM-ibid.
{\bf 65}, 1772 (1981).

\bibitem{browderpakvasa}
T.E. Browder and S. Pakvasa, University of Hawaii report, UH 511-814-95,
January 1995 (hep-ph/9501224).

\bibitem{dsnowmass}
I. Dunietz, in Proceedings of the Workshop on $B$ Physics at Hadron
Accelerators, Snowmass, Colorado, June 21 - July 2, 1993, p. 83, ed. by
Patricia~McBride and C.~Shekhar~Mishra.

\bibitem{muheim}
F.~Muheim, to be published in the proceedings of the DPF'94 conference,
University of New Mexico,
Albuquerque, New Mexico, August, 1994.  This reference did not include the
measured inclusive $\Xi_c$ yield in $\overline B$ decays that is not produced
by $\overline B \to \Xi_c \overline \Lambda_c X$.  Our calculation includes
this yield.

\bibitem{lipkin}
We thank Harry Lipkin for showing us the most elegant proof of that assertion
which follows.  Consider a complete set of states of $B_s$ decay modes governed
by $\overline b\rightarrow \overline c c \overline s$.  It can be chosen to
diagonalize the S-matrix.  Since the strong interactions conserve CP, a
complete set of states can be found where both the S-matrix is diagonal and
where each state has definite CP-parity. The existence of such a complete set
proves the assertion, because the total $b\rightarrow c\bar c s$ width for
$B_s$ decays is given by the sum of all the partial widths into CP eigenstates
and there are no cross-terms.

\bibitem{Dscleo}
T. Bergfeld et al., Cornell Report, CLEO CONF 94-9, 1994.

\bibitem{bcudbagan}
E. Bagan et al., TU Munchen Preprint TUM-T31-67/94/R (hep-ph/9408306), to
appear in Nucl. Phys.  {\bf B}.

\bibitem{earlier}
Q.~Hokim and X.Y.~Pham, Phys. Lett. {\bf B122}, 297 (1983);
Ann. Phys. {\bf 155}, 202 (1984).

\bibitem{pdg}
Particle Data Group, L.~Montanet et al., Phys. Rev. {\bf D50}, 1173 (1994).

\bibitem{slBR}
A.~Putzer, Heidelberg Report No. HD-IHEP-93-03 (1993).

\bibitem{butler}
J.N.~Butler and S.~Stone, contact persons for the Expression of Interest
 at Fermilab, EOI No.~2, May 1994.

\bibitem{qpnot1}
The quantity $|q/p|^2$ is predicted to be smaller than 1
by ${\cal O}(10^{-3} - 10^{-4})$, and $|p/q|^2$ larger than 1
by ``almost" the same amount. Thus the sum $|q/p|^2 + |p/q|^2$ is expected to
be 2
to much higher accuracy than ${\cal O} (10^{-3} - 10^{-4})$,
$$\left|\frac{q}{p}\right|^2 +\left|\frac{p}{q}\right|^2 = 2 + {\cal O} \left(
\frac{m^4_c}{m^4_t}\right) \;.$$
Dedicated future precision measurements may observe the tiny
$\Delta mt$-oscillations
in $\Gamma [g(t)]$ and $\Gamma [\overline g(t)]$. In contrast, such $\Delta
mt$-oscillations disappear to higher accuracy
in the sum $\Gamma [g(t)] +\Gamma [\overline g (t)]$.

\bibitem{ddgn}
C.O. Dib, I. Dunietz, F.J. Gilman, Y. Nir, Phys. Rev.
{\bf D41}, 1522 (1990).

\bibitem{rosnerdecayconstant}
J.L.~Rosner, Phys. Rev. {\bf D42}, 3732 (1990).

\bibitem{psikstar}
For a recent compilation see, for instance, T.E.~Browder,
K.~Honscheid and S.~Playfer, Cornell Report No.~CLNS
93/1261, in {\it $B$ Decays}, second edition, p. 158, ed.~S.~Stone,
World Scientific, 1994, and references therein.

\bibitem{kkps}
B. Kayser, M. Kuroda, R.D. Peccei and A.I. Sanda, Phys. Lett. {\bf
B237}, 508 (1990).

\bibitem{dqstl}
I. Dunietz, H. Quinn, A. Snyder, W. Toki and H.J. Lipkin, Phys.
Rev. {\bf D43}, 2193 (1991);
I. Dunietz, in {\it $B$ Decays}, second edition, p. 550, ed.~S.~Stone,
World Scientific, 1994.

\bibitem{bigirosner}
I.I. Bigi, D0/CDF lunch seminar given in 1993; J.L. Rosner,
private communication.

\bibitem{distinguish}
I.~Dunietz, Fermilab Report No.~FERMILAB--PUB--94/163--T, September 1994,
submitted to Phys. Rev. {\bf D}.

\bibitem{phitrigger}
Abreu et al. (DELPHI Collaboration), Ref.~\cite{delphi};
M.~Paulini, private communication;
J.~Incandela, E.~Kajfasz, R.~Snider and D.~Stuart, private communication.

\bibitem{ddw}
D. Du, I. Dunietz, Dan-di Wu, Phys. Rev. {\bf D34}, 3414 (1986);
Dunietz and Rosner, Ref.~\cite{dr};
Ya.I. Azimov, N.G. Uraltsev, V.A. Khoze, JETP Lett. {\bf 43}, 409 (1986).

\bibitem{stone}
See, for instance, S. Stone, Syracuse Report, HEPSY 94-5, September 1994.

\bibitem{glDphi}
M. Gronau and D. London, Phys.Lett. {\bf B253}, 483 (1991).

\bibitem{adk}
R. Aleksan, I. Dunietz and B. Kayser, Z.Phys. {\bf C54}, 653 (1992).

\bibitem{Dsk}
A feasibility study of extracting $\gamma$ from tagged, time-dependent studies
of $B_s \to D_s^\pm K^\mp$ has been conducted by E. Blucher, J. Cunningham, J.
Kroll, F.D. Snider, and P. Sphicas, in Proceedings of the Workshop on $B$
Physics at Hadron Accelerators, Snowmass, Colorado, June 21 - July 2, 1993, p.
309, ed. by Patricia~McBride and C.~Shekhar~Mishra.

\bibitem{aleksansumDsk}
R. Aleksan, A. Le Yaouanc, L. Oliver, O. P\`ene and J.-C.
Raynal, Orsay Report, LPTHE-ORSAY-94-03, 1994 (hep-ph/9407406).

\bibitem{atwoodsoni}
D. Atwood and A. Soni, SLAC Report, SLAC-PUB-6524, Jun. 1994 (hep-ph-9406391);
SLAC-PUB-6425, Jan. 1994 (hep-ph-9401347).

\bibitem{atwood}
D. Atwood, G. Eilam, M. Gronau and
A. Soni, SLAC Report, SLAC-PUB-6655, Sep. 1994 (hep-ph-9409229);
in preparation.

\bibitem{bsw}
M. Bauer, B. Stech, and M. Wirbel, Z. Phys. {\bf C34}, 103 (1987).

\bibitem{bijnens}
J. Bijnens and F. Hoogeveen, Phys. Lett. {\bf B283}, 434 (1992).

\bibitem{adsy}
D. Atwood, I. Dunietz, A. Soffer and H. Yamamoto,
Fermilab Report No.~FERMILAB--PUB--94/388--T, in preparation.

\bibitem{Wexchange}
Its relative importance could probably be assessed from the
CKM-favored transition of a non-strange $B_d\rightarrow D^-_s K^+$.
Because one deals here with strong interactions, rescattering effects, such as
$B_d \rightarrow D^- \pi^+ \rightarrow D^-_s K^+$, need to be disentangled.
Only then would one measure the square of the $W$-exchange amplitude.  The
$W$-exchange amplitude could be relatively more important to the
$B_s \rightarrow D^-_s K^+$ process, because the interference between the
$W$-exchange and the spectator contributes to the rate, in contrast to
the non-strange $B_d\rightarrow D^-_s K^+$ decay.

\bibitem{bss}
M. Bander, D. Silverman, and A. Soni, Phys. Rev. Lett. {\bf 43}, 242 (1979).

\bibitem{bjarlskog}
J. Bernabeu and C. Jarlskog, Zeit. Phys. {\bf C8}, 233 (1981).

\bibitem{Bdlifetime}
We  neglect the difference in the heavy and light $B_d$ lifetimes,
which are predicted to be at the one percent level.

\bibitem{bbhep}
F. Abe et al. (CDF Collaboration), Phys. Rev. Lett. {\bf 71}, 501 (1993);
Phys. Rev. Lett. {\bf 67}, 3351 (1991);
B. Adeva et al. (L3 Collaboration), Phys. Lett. {\bf B252}, 703 (1990);
D. Decamp et al. (ALEPH Collab.), Phys. Lett. {\bf B258}, 236 (1991);
H.C Albajar et al. (UA1 Collab.), Phys. Lett. {\bf B262}, 171 (1991).

\end{references}
\end{document}